\documentclass[journal]{vgtc} 
\vgtccategory{Research}

\newcommand\rev[1]{\textcolor{blue}}
\vgtcpapertype{application/design study}

\title{Investigating the Correlation Between Presence and Reaction Time in Mixed Reality}

\author{%
\authororcid{Yasra Chandio}{0000-0002-3436-6452}, \authororcid{Noman Bashir}{0000-0001-9304-910X}, \authororcid{Victoria Interrante}{0000-0002-3313-6663},
and \authororcid{Fatima M. Anwar}{0000-0001-7119-7232} 
}

\authorfooter{
  \item
  	Yasra Chandio is with University of Massachusetts Amherst.
  	E-mail: ychandio@umass.edu
  \item
  	Noman Bashir is with University of Massachusetts Amherst.
  	E-mail: nbashir@umass.edu.

  \item Victoria Interrante is with University of Minnesota Twin Cities.
  	E-mail: interran@umn.edu.
   
  \item Fatima M. Anwar is with University of Massachusetts Amherst.
  	E-mail: fanwar@umass.edu.

}

\abstract{%
Measuring presence is critical to improving user involvement and performance in Mixed Reality (MR). 
\new{\emph{Presence}, a crucial aspect of MR, is traditionally gauged using subjective questionnaires, leading to a lack of time-varying responses and susceptibility to user bias.} 
Inspired by the existing literature on the relationship between presence and human performance, \new{the proposed methodology systematically measures a user's reaction time to a visual stimulus as they interact within a manipulated MR environment.}
We explore the user reaction time as a quantity that can be easily measured using the systemic tools available in modern MR devices. We conducted an exploratory study (N=40) \new{with two experiments designed to alter the users' sense of presence by manipulating \emph{place illusion} and \emph{plausibility illusion}}. We found a significant correlation between presence scores and reaction times with a correlation coefficient -0.65, \new{suggesting that users with a higher sense of presence responded more swiftly to stimuli.}
\new{We develop a model that estimates a user's presence level using the reaction time values with high accuracy of up to 80\%. While our study suggests that reaction time can be used as a measure of presence, further investigation is needed to improve the accuracy of the model.}
}


\nocopyrightspace

\usepackage{xfrac}    
\usepackage{amsmath}
\usepackage{siunitx}
\usepackage{enumitem}
\usepackage{multirow}
\usepackage{graphicx}
\usepackage{subcaption}
\usepackage{flushend}
\usepackage[linesnumbered,algoruled,boxed,lined,noend]{algorithm2e}
\SetKwInOut{Parameter}{Parameters}

\newcommand\new[1]{\textcolor{black}{#1}}
\newcommand\revtwo[1]{\textcolor{black}{#1}}

\let\OLDthebibliography\thebibliography
\renewcommand\thebibliography[1]{
  \OLDthebibliography{#1}
  \setlength{\parskip}{0pt}
  \setlength{\itemsep}{0pt plus 0.3ex}
}

\begin{document}

\firstsection{Introduction}
\maketitle
\label{sec:introduction}
Mixed Reality (MR) is gaining importance in science, education, training, and entertainment, offering new ways of interaction and engagement with the real and virtual worlds. The technological advancements in MR tools have facilitated an enhanced sense of \emph{presence}, allowing users to behave within an MR environment as they would in the real world.
\emph{Presence} is typically described as the subjective experience of being in a simulated place or environment and the user's readiness to respond to virtually generated sensory data as if they were real~\cite{presence-1,presence-2,felton2022presence}.
This includes interacting naturally and appropriately with virtually generated sensory data. 
Just as in the real world, an individual should be able to bend down, grab an object on the floor in a virtual environment, feel its weight, and lift it if desired. This is achieved through the sense of one's body movement and position, which matches the sensory data presented in the virtual environment. High presence does not necessarily require high fidelity to physical reality but rather that individuals can behave as if the sensory data they are experiencing is real. 
This approach to measuring presence allows for observing and evaluating an individual's behavior in real and virtual environments.
A high \emph{presence} is desired in any simulated virtual environment, as it allows the user to engage in an immersive, realistic, and involved experience.

Many studies investigate the notion of \emph{presence} and describe the factors contributing to the sense of \emph{presence}. 
According to Witmer \& Singer ~\cite{ws}, Slater \& Steed~\cite{sus} and others~\cite{presence-1, presence-2, presence-measure-subjective-objective, felton2022presence}, there are two main aspects of \emph{presence}: \emph{place illusion} and \emph{plausibility illusion}. 
\emph{Place illusion} refers to the \emph{sense of being there}. 
In an MR environment, this corresponds to how the virtual content appears indistinguishable from the real world. 
\emph{Plausibility illusion} refers to the sensation that the observed events in a virtual environment occur. 
Users will feel involved in the environment when both \emph{place illusion} and \emph{plausibility illusion} occur. 
This involvement leads to users responding realistically to the environment, resulting in a greater sense of engagement in an MR environment. Immersion and participation are necessary to experience \emph{presence}~\cite{ws}.

A prerequisite to improving the \emph{presence} of a user is the ability to measure and quantify \emph{presence}. While conventional measures of presence have been defined for virtual environments that surround and isolate a user from the real world~\cite{presence-object-stevens2002putting}, we are measuring presence as the subjective experience that a particular object exists in a user’s environment, even when that object does not~\cite{arobject-questionaire-stevens2000sense}. Due to the subjective nature of \emph{presence}, the most popular method to measure \emph{presence} is the use of subjective questionnaires~\cite{ws,ipq,ipq-2, SSQ, questionaarie-measure-paper-grassini2020questionnaire}. Questionnaires ask users to self-report their sense of \emph{presence} by answering questions that attempt to assess \emph{presence}, usually after the user has left the virtual environment. The subjective questionnaires can be quickly administered, graded, and interpreted without affecting the user experience. However, questionnaires cannot measure the time-varying qualities of presence and can produce unstable, inconsistent, and irreproducible responses due to the prior experience of the participants~\cite{questionarie-prior-experience}. The well-known shortcomings of presence questionnaires have led researchers to explore alternative approaches to assessing presence, including behavioral responses such as postural response\cite{postural-response-freeman2000using}, hand and eye response~\cite{eye-response-wiederhold1998effects}, and startle response~\cite{old-presence-therory-starttle-response}, which are produced automatically, without conscious thought, thus avoiding user bias. However, the assessment of behavioral responses is susceptible to experimenter bias and is highly sensitive to environmental factors and content~\cite{behivor-quant-meaure,behivor-measure}. Various physiological responses can be measured to assess \emph{presence}, such as a change in heart rate~\cite{change-in-heart-rate2,change-in-heart-rate}, skin conductance~\cite{physiological-measure}, and body temperature~\cite{physiology-change}. However, physiological measures can also be noisy and unreliable, especially under non-stationary conditions, and may not capture differences in presence in situations of low emotional valence~\cite{physiological-measure}.

Given the state-of-the-art, there is a need for an approach to quantify \emph{presence} that is objective, quantitative, not consciously affected by the participant and/or experimenter, and could be used at runtime without interfering with the virtual experience. It should take advantage of existing interactions (and underline quantities) in the virtual scene and measure \emph{presence} through these interactions rather than making additional external interventions. One such underlying quantity is \emph{reaction time}. 
\emph{Reaction time} or response time refers to the time taken between when humans perceive something and when humans respond to it. \emph{Reaction time} is dictated by the cognitive ability to detect, process, and respond to a stimulus~\cite{responsetime-definition, reaction-time-cognition}, and can be easily measured using the systemic tools available in modern MR devices such as Microsoft HoloLens 2~\cite{hololens2}.

Our work investigates a fundamental question in MR: Would an individual experiencing more \emph{presence} systematically show faster \emph{reaction times}? 
If the answer is yes, we could use a systemic metric such as \emph{reaction time} to quantify \emph{presence} in a non-intrusive, objective, and unbiased manner. 
There is a large body of work investigating the relationship between \emph{presence} and human performance~\cite{sense-of-presence-barfield1993sense,presence-performance-maneuvrier-2020, 
presence-performance-review, presence-completion-time-cues, presence-performance-nature, presence-performance-measures}.
Natalia et al. showed a negative correlation between task completion time and \emph{presence} when the sense of \emph{presence} is altered by multisensory feedback~\cite{presence-completion-time-cues}. Matteo et al. show a negative correlation between performance and \emph{presence} when the sense of \emph{presence} is changed by varying the perceptual load~\cite{presence-performance-nature}. Maneuvrier et al. showed that presence promoted spatial cognition performance and that the presence-performance relationship was not mediated by other human factors~\cite{presence-performance-maneuvrier-2020}. Furthermore, human performance is often used as an argument for the good predictive validity of questionnaires~\cite{questionaarie-measure-paper-grassini2020questionnaire}. 

To understand the relationship between \emph{presence} and \emph{reaction time}, we conducted a study in which we varied the \emph{presence} of users by manipulating \emph{place illusion} and \emph{plausibility illusion} while they were interacting with an MR environment. We designed two sets of experiments. In one set, we only manipulated the appearance of the virtual object, and in the other set, we manipulated a non-task-relevant behavior of the virtual object. All other aspects of the experiments, such as the interaction mechanism, frequency of interactions, and physical environments, were kept the same. We systematically measured the \emph{reaction time} of users in response to a visual stimulus. Our post-experience questionnaires show a significant change in presence in each experiment between the manipulation conditions.
Similarly, we observed a significant change in user \emph{reaction time} as the sense of \emph{presence} changed.
Our analysis shows a correlation between \emph{presence} and \emph{reaction time}. 

In our attempt to understand the relationship between \emph{presence} and \emph{reaction time}, this work makes the following contributions. 

\noindent \textbf{Contribution 1:} We propose the use of \emph{reaction time} of a user as a measure of \emph{presence}. We also develop a non-intrusive, systemic approach to measuring the user \emph{reaction time} that relies on existing interactions in MR environments. 

\noindent \textbf{Contribution 2:} We devise experiments that alter the sense of \emph{presence} of a user by manipulating \emph{place illusion} and \emph{plausibility-illusion}. In designing experiments, we control for other factors that are known to impact user performance. While we use the experiments to demonstrate change in presence, we also demonstrate, as a byproduct, that presence questionnaires typically used in fully virtual environments can also be used in MR environments.

\noindent \textbf{Contribution 3:} We conduct an exploratory lab study ($N$ = 40) that demonstrates a negative correlation between the sense of \emph{presence} and the user \emph{reaction time} when responding to a visual stimulus. 

\noindent
\new{
\textbf{Contribution 4:} We develop a model that estimates a user's presence level using the reaction time values as input. Our evaluations demonstrate that model has high accuracy (up to 80\%), which can be further improved with data from a larger number of users. 
}


\section{Background and Related Work}
\label{sec:background}
In this section, we define the relevant terminology, discuss the existing work on the concept of presence in MR, and present existing methods for quantifying presence.

\subsection{Terminology}
\begin{figure}[t]
    \centering
    \includegraphics[width=\linewidth]{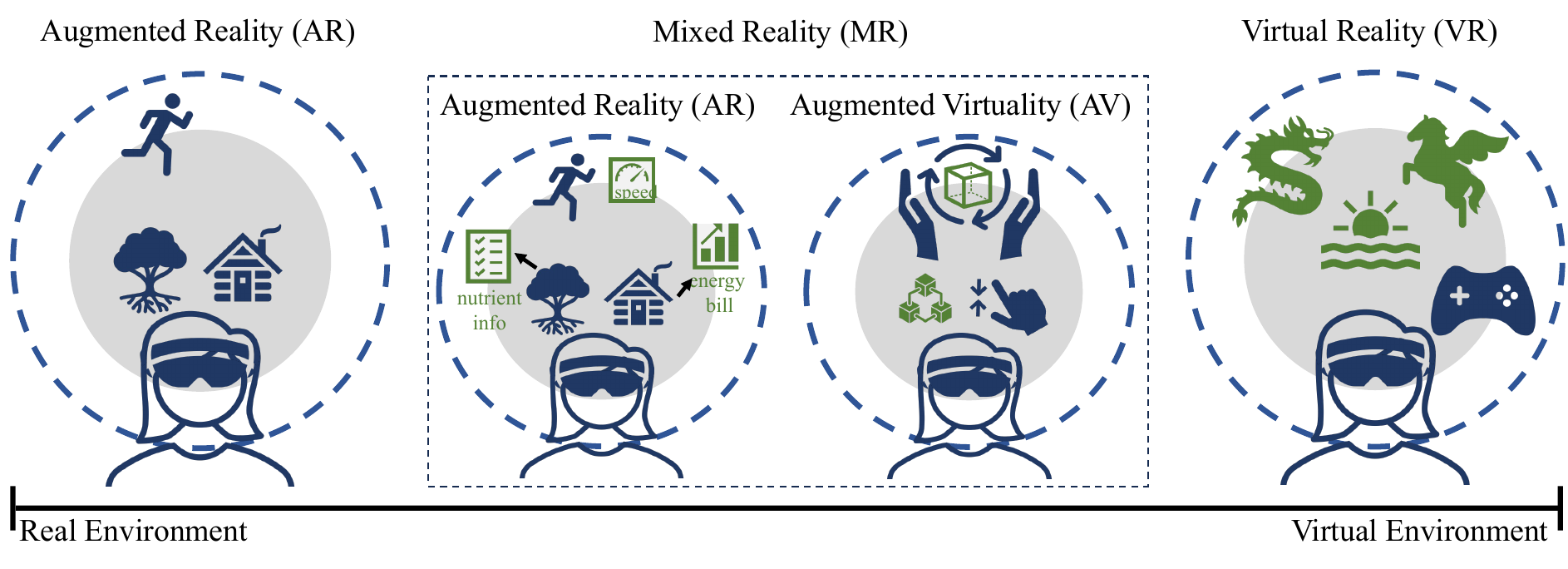} 
    \vspace{-0.48cm}
    \caption{\new{\emph{\textbf{The Reality-Virtuality Continuum~\cite{milgram1995augmented}. Virtual objects are colored green, and real-world objects are colored blue.}}}}
    \vspace{-0.65cm}
    \label{fig:continuum}
\end{figure}

Different forms of \emph{reality} depend on how much of the physical world is part of the user's experience and how the user interacts with the virtual objects in the scene, as shown in Figure~\ref{fig:continuum}.
\new{
Defining MR remains challenging, with no universally agreed-upon comprehensive definition~\cite{what-is-mr-chi}. Virtual Reality (VR) immerses users in a wholly digital realm, while Augmented Reality (AR) superimposes digital elements into our real world. 
\revtwo{Augmented Virtuality (AV) is an immersive experience, complete or partial, with added elements of 'reality' such as video or texture mapping.} 
MR is an umbrella term that encompasses both AR and AV. Our MR concept leans towards AR on Milgram's reality-virtuality spectrum, where users interact primarily with virtual objects while being able to see the real world around them.  
MR represents a spatial alignment between the real and virtual worlds, allowing users to interact with and manipulate both real and virtual environments. We will use MR as a blanket term throughout the paper~\cite{milgram1995augmented}.
}

\subsection{Factors Affecting Presence}
\emph{Presence} is a phenomenon of awareness based on the interaction between sensory stimulation, environmental factors that encourage involvement and allow immersion, and internal tendencies to become involved and interact with virtual objects~\cite{ws}. Sheridan~\cite{sheridan1992musings} laid the foundation for determining the underlying presence factors, such as sensory information, sensor control, and motor control. Slater and Wilbur~\cite{Place-illusion-plausibility, wilburandSlater1997AFF} expanded on Sheridan's work to determine major factors affecting user presence. According to them, two main factors contribute the most to user presence.
(1) \emph{Place illusion} refers to the appearance of a virtual environment (or virtual object in the case of AR/MR).
It can be affected by the realism of the virtual content, the consistency of the view between the headset and the direct view from the users' eyes (including displacement or latency issues), the lack of haptic feedback, and the awareness of the headset.
(2) \emph{Plausibility illusion} refers to the behavior of a virtual environment (or virtual object in the case of AR/MR). 
It can be affected by scene elements not obeying the laws of physics, cause-effect relationships not coupled as expected, actions that do not have the expected outcome, and events in the virtual environment not conforming to familiar expectations.
It is argued that when both \emph{place illusion} and \emph{plausibility illusion} occur, users will feel involved and respond realistically to the environment, which will lead users to experience greater engagement in a virtual environment~\cite{ws}.

\subsection{Presence Measures}
\label{sec:presence-measures}
Previous research highlights the importance of presence
as an outcome of virtual environments (AR/VR/MR)~\cite{sheridan1992musings, ws,  presence-1, presence-2}. 
The most commonly accepted method of evaluating \emph{presence} is the self-report questionnaire~\cite{questionarrie-in-realities-usoh2000using}. Typically,
presence questionnaires are used after participants engage in an AR/VR/MR environment, making them post-experience questionnaires. Over time, previous research has developed, refined, and validated questionnaires for measuring presence. 
Bystorm et al.~\cite{sense-of-presence-bystrom1999conceptual} 
proposed an integrated theoretical framework for studying presence that includes aspects related to task performance. In their work, they used two questionnaires to measure interaction fidelity.

The Witmer \& Singer Presence Questionnaire (PQ) is one of the most widely used~\cite{ws} to assess involvement/control, naturalness, and interface quality. Other popular post-experience questionnaires include the Igroup Presence Questionnaire (IPQ)~\cite{ipq,ipq-2} and the Slater-Usoh-Steed Questionnaire (SUS)~\cite{sus}. The PQ questions are based on the following factors: the ability to control the environment and "naturalness" of control over the environment, coherence and consistency of information from different senses, and distractions a participant may experience in a virtual environment. They also consider environment realism and meaningfulness, as well as the sense of disorientation when returning to the real world. 
On the other hand, the SUS and IPQ questions are based on three factors: the sense of physically being in the virtual environment, the extent to which the virtual environment feels real, and the extent to which the participant feels involved. 
There are additional questionnaires that measure various aspects of realism relating to the virtual scene, such as the ITC-Sense of Presence Inventory~\cite{ITC-questionarrielessiter2001cross}, the Kim and Biocca questionnaire~\cite{kim1997telepresence}, the Object presence questionnaire~\cite{presence-object-stevens2002putting}, the reality judgment and presence questionnaire~\cite{banos2000presence}, the Swedish viewer-user presence questionnaire (SVUP)~\cite{larsson2001actor}, and others~\cite{sense-of-presence-barfield1993sense,  nichols2000measurement, choquestionarrie2003dichotomy}. 
However, these questionnaires do not aim to measure the time-varying qualities of presence. They can produce unstable, inconsistent, and unreproducible responses, and are susceptible to user bias. To address the issues of temporal information issue, Schwind et al.~\cite{schwind-pinky} used integrated questionnaires to measure presence. Integrating questionnaires directly into the virtual experience has been explored to assess the virtual experience as the user is going through it~\cite{not-my-hands, schwind-pinky, in-VR-questionarrie}. However, the problem of defining a clear baseline remains. 

Other measures are continuous assessments~\cite{time-varying-questionarie-input,presence-measure-subjective-objective}, psychophysical measures like cross-modality matching, free-modulus magnitude estimation, and paired comparisons. Subjective qualitative methods include autoconfrontation~\cite{autoconfrontation-method-retaux2003presence}, focus group exploration~\cite{freeman2000focus}, interaction analysis~\cite{spagnolli2003ethnographic}, free format self-reports~\cite{turner2003re}, and repertory grid technique~\cite{steed2003experiences}. There are also several subjective corroborative measures to evaluate presence indirectly, such as Break In Presence (BIP)~\cite{break-in-presence}, duration estimations by users~\cite{ijsselsteijn2001duration}, attention awareness~\cite{darken1999quantitative}, and simulator sickness questionnaire~\cite{SSQ}. Most of these methods ask participants to complete questionnaires that often contribute to the phenomenon of break-in presence~\cite{break-in-presence} and are prone to participants' bias.

Objective measures are mostly captured by evaluating the behavioral and physiological responses of the users and are often used as corroborative measures. Popular physiological measures are cardiovascular measures~\cite{change-in-heart-rate}, 
skin measures~\cite{physiological-measure},
ocular measures~\cite{laarni2003using},
and facial electromyography~\cite{huang1999presence}. Presence is also known to be measured through neural correlates like Electroencephalogram (EEG) and Functional Magnetic Resonance Imaging (fMRI)~\cite{hoffman2003illusion}.  
Physiological measures can reduce user bias but may also be prone to inaccuracy and unreliability~\cite{physiological-measure}. 
Some commonly used behavioral measures are based on assessing facial expression~\cite{behivor-measure}, postural~\cite{postural-response-freeman2000using}, 
startle~\cite{old-presence-therory-starttle-response}, reflex and social responses, and pointing conflicting cues~\cite{behivor-quant-meaure}. 

\subsection{Presence and Performance}
Many studies have discussed the association of presence in the context of task performance in a virtual environment ~\cite{lombard1997heart, sus, sense-of-presence-barfield1993sense, basdogan2000experimental, perfromance-presence, ijsselsteijn2004presence, human-performance-1}, we discuss only a few of them here.
Slater et al.~\cite{perfromance-presence} conducted experiments to assess the influence of presence on performance while the participants learned to play three-dimensional chess.  
They noted that presence refers to the behavioral and psychological responses of people. 
Similarly, Barfield et al.~\cite{sense-of-presence-barfield1993sense} suggested that task performance measures can be used as objective corroborative indicators of presence. A few such methods are task completion time and error rate~\cite{basdogan2000experimental}, the number of actions~\cite{perfromance-presence}, and secondary task performance~\cite{ijsselsteijn2004presence}. Though it is generally assumed that higher levels of presence are associated with better task performance~\cite{human-performance-1},
the exact causal link between presence and task performance is unclear. 

Slater et al.~\cite{perfromance-presence} explored the relationship between presence and performance. 
While keeping other factors such as relevant background knowledge and users' ability the same, results suggested that increasing presence by increasing the richness of the virtual environment improved task performance. 
The study also found that reported presence was higher for egocentric than exocentric immersion, but a causal relationship between presence and task performance was not established~\cite{wilburandSlater1997AFF}.
It is also noted that motor behavior is strongly influenced by perceptual uncertainty and the expected consequences of actions~\cite{perceptual-uncertainity} that can affect user's characteristics, such as ability and motivation, that will influence task performance~\cite{heeter2003reflections}. 
IJsselsteijn et al.~\cite{ijsselsteijn2004presence} noted that it is reasonable to assume that several characteristics of a virtual environment will similarly influence presence and task performance. 
They further expanded that performance on a secondary task can serve as a measure of the amount of effort and attention allocated to the primary task. 
The more effort is dedicated to the primary task, the more performance on the secondary task will decrease. A similar argument can be made in the case of presence: if more attention is allocated to the mediated virtual environment, performance on a secondary task will decrease. 
Szczurowski and Smith made it reasonable to assume that the nature of presence is subconscious~\cite{reaction-time-szczurowski2017measuring}. They argue that if the presence exists outside the subjective feeling domain, it’s unlikely to be a conscious process. No one must remind themselves about staying present in the real world. It’s also unlikely that it is possible to force yourself into feeling present in a virtual environment.
IJsselsteijn, Szczurowski, Smith, and others~\cite{reaction-time-cognition, reaction-time-correaltion, reaction-time-szczurowski2017measuring, musings-after25yers-barfield2016musings} concluded that reaction time or error rate could be used as task performance measures for presence evaluation.

 Therefore, we make a case for measuring subjective feelings of presence with an objective, reproducible approach and producing a stable response without interference in the virtual scene. 
 Despite being aware of problems associated with a subjective measurement method, we are deciding to use questionnaires as the baseline for our current study to test our hypotheses about relations between different constituents of \emph{presence} and users' \emph{reaction time}.

\subsection{Presence in Less-immersive AR/MR Environments}
Presence governs aspects of a user's autonomic responses and behavior in a virtual environment, whereas immersion refers to a quantifiable description of a technology~\cite{slater-usoh-steed1995taking}. Wilbur and Slater argue that the degree of immersion can be objectively assessed as the characteristics of a technology~\cite{wilburandSlater1997AFF}, and has dimensions such as the extent to which a display system can deliver an inclusive, extensive, surrounding and vivid illusion of a virtual environment to a user~\cite{vivideness-def-steuer1992defining}.
 AR/MR elicits a different sense of presence: ``It is here'' presence ~\cite{ar-presence-at-heart-of-it-all-lombard1997heart}. 
Although AR/VR/MR is very different from a technical perspective, a common feature they share is that virtual objects exist in a curated environment: real (in the case of AR and MR) or virtual (in the case of VR).
 Therefore, the common approach to measuring presence in various virtual environments, from the least immersive environment (AR) to the fully immersive environment (VR), questions whether one has a sense of being in or interacting with the virtual environment. 
 Although there have been attempts to develop presence methods~\cite{ar-presence-gandy2010experiences, presence-non-immersive,presence-object-stevens2002putting} and measurement tools~\cite{AR-presence-Regenbrecht2021MeasuringPI,arobject-questionaire-stevens2000sense} exclusively for AR/MR, these tools only measure factors that may influence presence, rather than directly measuring the subjective sense of presence itself~\cite{ presence-mr}.
 
While conventionally, presence has been defined for virtual environments that surround and isolate a user from the real world~\cite{presence-object-stevens2002putting}, Slater et al. \cite{questionarrie-in-realities-usoh2000using} conducted a study on VR questionnaires (PQ and SUS) and concluded that the questionnaires that are developed for VR can still be useful when all users experience the same type of environment even if the environment is not fully immersive (AR/MR). They also concluded that the utility of questionnaires might be doubtful for comparing experiences across environments - such as immersive virtual compared to real, or desktop compared to immersive virtual, or a real environment with virtual objects to a fully virtual environment. Presence questionnaires are often utilized in research to explore the subjective experience of presence rather than the link between perceived presence and aspects of technology; therefore, they can be employed anywhere on the virtuality continuum in technological or real-world contexts. To this end, in this study, we refer to \emph{presence} as the subjective experience that a particular object exists in a user’s environment, even when that object does not~\cite{arobject-questionaire-stevens2000sense}. This definition is more appropriate for assessing non-immersive displays such as AR/MR headsets~\cite{presence-non-immersive}. We use this definition of \emph{presence} for the rest of the paper. 

\subsection{Taxonomy of a virtual scene}

As described by Wilbur and Slater~\cite{wilburandSlater1997AFF}, the place illusion (appearance realism) and the plausibility illusion (behavioral realism) are the main aspects of any virtual experience in any alternative reality medium.

 \noindent \textbf{\textit{Virtual scene}} represents the semantics of the virtual environment in three dimensions (3D) placed within the real environment. 
 
    \noindent \textbf{\textit{Event}} An event
    happens in a computer system. For instance, adding or removing a virtual object from a virtual scene is considered an event.
    
    \noindent \textbf{\textit{Task}} refers to an observable activity with a start and an endpoint. In MR, tasks will be aligned with the start or end of an event, depending on the semantics of the virtual scene.
    
    \noindent \textbf{\textit{Interaction}} is defined as performing a physical action to perform the task in a virtual scene. 
    
    \noindent \textbf{\textit{Cue}} is defined as the signal (visual or auditory) that is sent to the participant to initiate a task.
    
    \noindent \textbf{\textit{Feedback}} is the visual or auditory confirmation sent to the participant that the task is completed.

To create a virtual scene with context, immersion, and interaction, we need to craft our experiments so that participants feel engaged in the virtual world.
However, the scene should neither be too complex that a participant's cognitive load is consumed in understanding the scene nor should the scene be too simple that the participant feels disengaged~\cite{slater-usoh-steed1995taking,ijsselsteijn2004presence}. 
For example, we do not want a scene with Warcraft heavy-load games or a simple box with no semantic value to the participant. We need to create a balance between a semantically too complex and a simple virtual scene. 
The same applies to events, tasks, cues, interactions, and feedback from the virtual scene.

\section{Approach}
\label{sec:design}
\new{To validate \emph{reaction time} as a measure of \emph{presence}, we investigate the \emph{correlation between presence and reaction time} in MR.
We reference Insko et al.'s ~\cite{presence-measure-subjective-objective} criteria for a useful measure: \textit{sensitivity} (to detect different levels of 'presence'), \textit{reliability} (providing repeatable results), \textit{validity} (correlating with existing 'presence' measures), and \textit{objectivity} (free from participant's and experimenter's bias). Accordingly, we navigate the following design challenges (DCs):}

\noindent  \new{\textbf{DC1:} How can we induce different feelings of \emph{presence}? (sensitivity)}

\noindent \new{\textbf{DC2:} How do we \emph{minimize confounding variables} while varying feelings of \emph{presence}? (reliability)}

\noindent \new{\textbf{DC3:} How do we \emph{establish a baseline} measure of presence and what should that baseline be? (validity)}

\noindent \new{\textbf{DC4:} What user \emph{interaction mechanism} should we use to assess reaction times? (objectivity)}

\subsection{DC1 - Varying presence}
We need an empirical setup that can measure presence in real-time while depicting a practical scenario for measuring varying feelings of presence. But first, we need to understand the main aspects of presence and what dimensions could be measured in those distinct but overlapping aspects. We describe two main aspects (place illusion and plausibility illusion) of presence in~\S\ref{sec:background}. Since we are using Mixed Reality (MR) as our experiment medium, place illusion and plausibility illusion need to be refined. In MR, measuring the place illusion could mean to what extent the virtual object appears indistinguishable from reality. The sense of plausibility can be described when users select a dominant space as the reference frame. Then virtual objects in real space or real objects in a virtual space would be perceived as plausible if the object behaves coherently to the dominantly perceived space as noted in place illusion. For example, plausibility would be lessened if gravitational forces were applied horizontally rather than vertically. 

In summary, place illusion refers to the elements related to the \emph{appearance} of the environment. In contrast, plausibility illusion refers to the elements related to the \emph{behavior} of the objects in the environment. Therefore, we suggest that to vary the feelings of presence, we could manipulate \emph{appearance} 
and manipulate the \emph{behavior} of the object in the scene. 
\new{
As Slater describes, presence is affected by both realism and plausibility. To establish the relationship between response time and presence, irrespective of why the presence changes, we altered realism and plausibility to induce various levels of presence. Other factors can also impact presence, and we plan to explore their effect in the future.
}

\subsection{DC2 - Controlling confounding variables}
\label{sec:experiment}
Various options exist for manipulating objects' appearance and behavior in the virtual environment. 
We must pick scenes and manipulations carefully to satisfy the reliability criteria. We have identified two constraints that will help limit the introduction of external variables and maintain symmetry across experiments and users. 

\begin{itemize}[leftmargin=*, topsep=0pt,itemsep=-0.8ex]
\item \textbf{Constraint 1:} Manipulations should not affect (increase or decrease) the overall complexity of the scene, including tasks, events, and interactions. (\textbf{simplicity} in scene)
\item \textbf{Constraint 2:} Manipulations should also be free from additional confounding variables. Confounding variables are the extra variables that affect the actual relationship between the variables under study~\cite{confounding-variable}. These output variables are \emph{presence} and \emph{reaction time}. (\textbf{symmetry} in the scene and users)
\end{itemize}

To satisfy the simplicity constraint, we will avoid unnecessary details in the scene (features mentioned above), and we are enforcing consistency in the non-manipulated conditions of our experiment. 
\new{The selection of simple scenes is a conscious design choice to control the effect of confounding variables. In a complex scene, it can be hard to isolate the effect of various factors on presence.} These experimental conditions are the placement and duration of the visual stimulus, time and appearance of the cue, time and appearance of the feedback, number of tasks, duration of the experiment, and type of interaction. We also suggest that to keep the scene simple, we should pick a scene familiar to most college students (intended participants pool) and have only one virtual object in the scene at a time. However, we want to keep the participant engaged throughout the experiment, and we need a scene with some familiar semantics. Similarly, the manipulations that will vary the presence should be subtle and change only one scene to isolate the effect on our output variables. In~\cite{abstract-real}, authors have proposed manipulating the scene's appearance
by manipulating the visual fidelity. We vary presence by making the scene appear \emph{realistic} in one scenario and \emph{abstract} in another. Similarly, as shown in~\cite{break-in-presence}, plausibility illusion can be affected by challenging the physical laws in a scene. 
In behavior manipulation, we make the behavior of a scene naturally plausible in one scenario and implausible in another. We will maintain a controlled physical environment to minimize interference and satisfy symmetry constraints. 
We keep the same room, lighting, seating arrangement, study session duration, breaks, and order of the experiment across users and experiments. 

To this end, we divide our experiment into two sets. In the first set, we vary the sense of presence by 
manipulating the \emph{appearance} of the virtual object (place illusion). In the second set, we vary the sense of presence by manipulating the \emph{behavior} of the virtual object (plausibility illusion). Each experiment will contain two blocks of trials: a \emph{control} block and a \emph{manipulated} block. In the \emph{control} block, the scene contains virtual objects with a natural appearance and plausible behavior. However, in the \emph{manipulated} block, the scene contains virtual objects whose appearance is manipulated to be visually unnatural (unrealistic) or in which the virtual objects exhibit implausible behavior.

\begin{figure*}
    \centering
    \includegraphics[width=\textwidth]{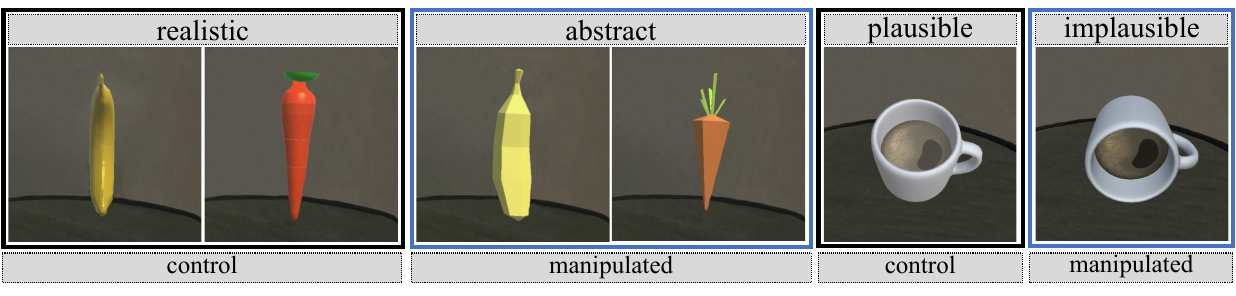} 
    \vspace{-0.55cm}
    \caption{\textbf{\emph{Realistic, abstract, plausible, and implausible virtual objects used in the experiments. }}}
    \vspace{-0.6cm}
    \label{fig:main-illusion}
\end{figure*}

\subsection{DC3 - Establishing a baseline}
\label{sec:questionnaire}
 
Since presence itself has a subjective nature, it is logical that we also establish our baseline with a subjective measure. To understand the level of presence, we use the questionnaire as our baseline, which researchers commonly accept as the standard measure. This helps us set a starting point for our study~\cite{questionnaire-chi-20, break-in-presence, abstract-real}. Subjective questionnaires have been the standard measure of presence for many years. They are sensitive enough to find differences in presence when used to examine the difference between two visually similar fidelities~\cite{presence-measure-subjective-objective}. The post-experience questionnaire provides scores that reflect the level of perceived presence in the scene. 
\new{
While subjective questionnaires have limitations, as discussed in \S\ref{sec:presence-measures}, they are currently the only widely accepted method of quantifying presence. 
}
We measure various feelings of presence with the three most-cited and widely used questionnaires that measure presence. We use the Igroup Presence Questionnaire (IPQ)~\cite{ipq,ipq-2}, Slater-Usoh-Steed (SUS)~\cite{sus}, and Witmer \& Singer (PQ)~\cite{ws}.

\subsection{DC4 - Selecting an interaction mechanism} 
We need a task that engages the participant and helps us measure our output variable, \emph{reaction time}. 
Through interactions, the participant can physically be involved in the scene. 
In designing the interaction, we want to avoid imposing differential barriers to task completion, such as placing a button at a height that is easier to reach for some participants than others.
Similarly, we want to avoid designing unnecessarily complex interactions that could cause task performance to vary unpredictably, independent of the targeted manipulations. 
Additionally, we ensure that external interventions do not affect the reaction time.
Therefore, we need to leverage existing elements in the taxonomy of the scene (cue, interactions, and feedback). 

\new{
To solve these challenges, we employ HoloLens's ``air tap'' gesture~\cite{air-tap} as our interaction mechanism. It is precise, and its ease of use is independent of a typical participant's height, reaching range, or any other physical attribute. The air tap gesture requires specific movements performed in a particular order. According to the instructions for air tap in the HoloLens 2 manual~\cite{hololens-air-tap}, the user needs to "hold your hand straight out in front of you in a loose fist, point your index finger straight up toward the ceiling, tap your finger down, and then quickly raise it back up again." This specific sequence of movements reduces the probability of mistriggering due to random motions. Additionally, in the scene depicted in Figure~\ref{fig:air-tap}, there is no movement other than the participant performing an air tap in response to a cue. Therefore, the closed position of the participant's fingers in the image is not an error and does not pose a potential for inaccurate gesture recognition.
}

We designed a task that supports participants in knowing what to do (cue), knowing that the system is working (interaction/air tap), and knowing if their action was understood by the system (feedback).
In the \emph{realistic} vs. \emph{abstract} scenario, 
the appearance of the object in the scene is the cue for the user to take action (air tap), and the disappearance of the object from the scene is the feedback to the user that their action was successful.
In the \emph{plausible} vs. \emph{implausible} scenario, the cessation of change in the height of the coffee in the cup cues the user to initiate their action. The coffee cup changes color to provide feedback to the user about the success of their action (details in~\S\ref{sec:pilot-feedback}).

Next, we formally define the reaction time (${reaction}_i$) in our study for the trial $i$ as $({interaction}_i - {cue}_i)$. 
where trial $i$ refers to one task iteration by the participant (cue, air tap, feedback).  
${interaction}_i$ is the time when the air tap is recorded in trial $i$, and ${cue}_i$ is the time of onset of the task cue in trial $i$.

\section{User Study}
\label{sec:study}

In this section, we detail the study measures, participants, equipment, and procedure for the user study. 

\subsection{Participants}
We recruited 40 participants (23 male-identifying, 16 female-identifying, and one non-binary identifying) with a mean age of 26.6 years (standard deviation of 5.5 years). 
All participants volunteered and provided written informed consent. They received \$25 for their participation. All but one participant had a technical background in computer science or engineering. All the participants had normal or corrected normal vision with contact lenses or glasses. 
Twelve participants had 1 to 4 days per week of MR experience, 24 participants had less than 1 hour per week of experience, and nine had never experienced MR before. 
Only 12 participants had used Hololens 2 before the study.
The study was granted ethics clearance according to the ethics and privacy regulations of our Institutional Review Board (IRB). 

\subsection{Material} 

\new{The study utilized  
an ergonomic, untethered, self-contained holographic device, Hololens 2~\cite{hololens2} equipped with a second-generation Holographic Processing Unit (HPU) for real-time computer vision and a Qualcomm Snapdragon 850 CPU for running applications. 
The virtual scenes were developed via Unity 3D (10.0.19362.0) game development engine 
with API for the Universal Windows Platform on Windows 10 PC.
}

\new{The Hololens 2 accepts eye, spatial, and hand-tracking inputs with a field of view of $43^{\circ}$ horizontal, $29 ^{\circ}$ vertical, and $52^{\circ}$ diagonal. Its dual see-through displays have a resolution of $1440 \times 936$ pixels each, a 60Hz refresh rate, and a tinted visor to minimize environmental light interference. We chose the HoloLens 2 for our experiment because its see-through setup allows the user a direct view of the real world. The other MR headsets that leverage video to show the physical world cannot be used for safety-critical applications like surgery.}

\begin{figure}[t]
    \centering
    \includegraphics[width=0.5\linewidth]{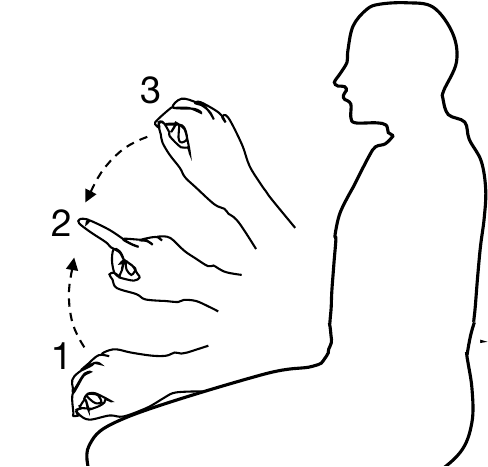} 
    \vspace{-0.05cm}
    \caption{\emph{\textbf{Illustration of the experimental task using air tap.}}}
    \vspace{-0.55cm}
    \label{fig:air-tap}
\end{figure}

\subsection{Variables}
In our study design, we change two variables to test our hypothesis, but we only manipulate one variable at a time. 
We chose \emph{appearance} and \emph{behavior} of virtual objects as our variables in the first and second experiment sets, respectively. 

\subsubsection*{\textbf{Experiment Set 1: Realistic vs. Abstract}}
In the \emph{control} trials, all virtual objects in the scene will have a realistic appearance (textured and natural) and plausible behavior. 
In the rest of the paper, we refer to this block of trials as \emph{realistic}. 
In the \emph{manipulated} trials, all virtual objects will depict plausible natural behavior, but the appearance would be abstract (untextured and geometric). 
We refer to this block of trials as \emph{abstract} in the remainder of the paper.

In the control trials, all of the virtual objects in the scene have a realistic appearance (textured and natural) and plausible behavior. In the rest of the paper, we refer to
this block of trials as \emph{realistic}. In the manipulated trials, all virtual objects depict plausible natural behavior, 
but their appearance is abstract (untextured and geometric). We refer to this block of trials as \emph{abstract} in the remainder of the paper. 
For the \emph{realistic} vs. \emph{abstract} trials, we chose a scene that contains textured virtual objects. To satisfy our simplicity constraint, we modified the popular Fruit Ninja game~\cite{fruit-ninja}. To mimic the semantics of a multi-object environment, we use two objects that are similar in shape and size but different in textural properties. As we wanted a simple scene and one virtual object at a time, we made the virtual objects appear in the scene one after the other. Ultimately, we used one fruit (banana) and one vegetable (carrot) for this experiment. We removed complex interactions (slicing) and additional cognitive loads (scores) from the original game. In the \emph{realistic} condition, we make the carrots and bananas appear as natural as possible in terms of color, texture, and shape. In the \emph{abstract} condition, we render the carrots and bananas with muted colors, without texture, and with geometrically blocky shapes, as shown in Figure~\ref{fig:main-illusion}.

\begin{figure*}[t]
    \centering
    \includegraphics[width=\textwidth]{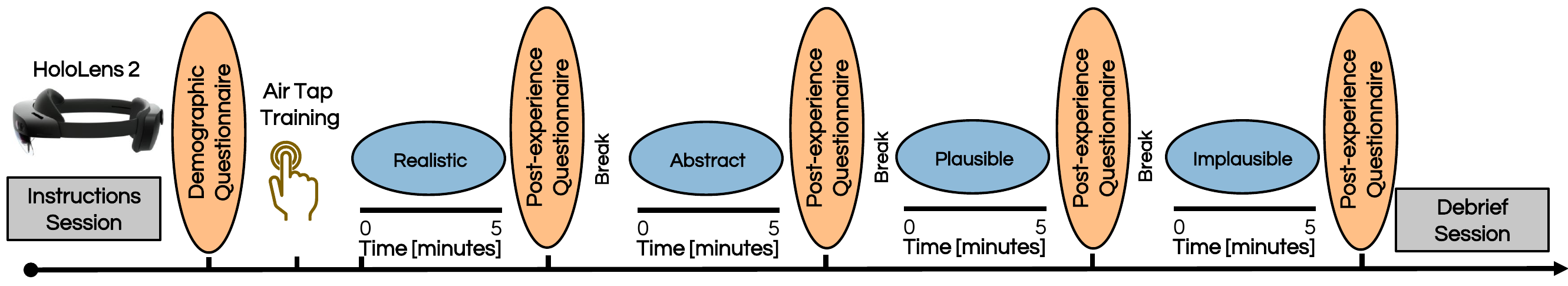} 
    \vspace{-0.65cm}
    \caption{\emph{\textbf{User study timeline consisting of pre-and post-questionnaire and four blocks across the two experiments.}}}
    \vspace{-0.6cm}
    \label{fig:study-flow}
\end{figure*}

\subsubsection*{\textbf{Experiment Set 2: Plausible vs. Implausible}}
In the \emph{control} condition, the virtual objects in the scene have a natural appearance and plausible behavior. We refer to this block of trials as \emph{plausible} in the remainder of the paper. In the \emph{manipulated} condition, the virtual objects retain a natural appearance but behave implausibly from real-world standards (disobeying laws of physics). In the remainder of the paper, we refer to this block of trials as \emph{implausible}.

For the \emph{plausible} vs. \emph{implausible} scenario, we chose a scene with elements that have physical constraints.  We modified the popular Coffee Stack game~\cite{coffe-stack} with a coffee mug on a surface that periodically fills up with coffee.
Like Fruit Ninja, we have removed complex interactions and cognitive loads from this game. 
In the \emph{plausible} scenario, the empty white porcelain coffee mug is upright, and the coffee pours into the mug periodically until the mug is full. 
Then, a new mug appears, and this cycle continues to repeat. 
In the \emph{implausible} scenario, the empty white porcelain coffee mug is in a tilted position on the surface, and coffee continues to pour into and fill the mug even though gravity and volumetric constraints should not allow this (see Figure~\ref{fig:main-illusion}).

\subsection{Experimental Task}
The experimental task in both sets can be divided into three sub-tasks, as shown in Figure~\ref{fig:air-tap}. The experiment objects are placed in the participant's field of view as a prerequisite. The first step for participants is to view the object with their hand in a neutral position without raising their elbow, as illustrated in Figure~\ref{fig:air-tap}(1) and captured in Figure~\ref{fig:main-illusion}. The second step for the participants is to lift their index finger and react upon cue by air-tapping the virtual objects, as shown in Figure~\ref{fig:air-tap}(2). 
Finally, the third step for participants is to return their hands to the neutral position after seeing the feedback.
\new{To avoid double triggers, we instructed participants to perform a single air tap in response to a cue.
}

\subsection{Measures}
Presence scores for the questionnaires were obtained using 39 items (6 SUS, 14 IPQ, 19 WS) on a 7-point scale. We did not modify any of the questions. The reaction time is recorded by our software on HoloLens 2.
On average, we collected 50 reaction time measurements per block of trials, totaling 200 data points per participant for the four blocks. 
The reaction time is measured in milliseconds ($ms$). 

\subsection{Pilot Study}
Before starting the formal study, we conducted a pilot study with two participants to tune the parameters to minimize environmental variables and maximize reliability and objectivity. 
We use a talk-aloud protocol and ask participants questions to align the experiment for general comfort and ease for the participants but not to bias the experiment for a specific set of users. 
We tuned the following session-specific parameters.

\noindent{\textbf{\emph{Participant position.}}}
We experimented with standing, walking, and sitting positions. The participants reported feeling tired while standing. While walking, they reported that the scene kept changing around them, necessitating extra attention to locate the virtual object. Participants reported the sitting position as the most comfortable position for the experiment. 
We asked the participant to sit on a chair with relaxed shoulders, an arm on the lap or armrest, and feet flat on the floor. The chair with armrest and backrest was reported to be the most comfortable with air tap interaction~\cite{air-tap}. 

\noindent{\textbf{\emph{Room lighting.}}} 
Lighting affects a user's ability to see the virtual environment and its objects~\cite{lighting-ost}.
It also affects the rendering of virtual objects and user interaction with those objects, as they rely on the tracking module of Hololens 2. 
For optimal visuals through Hololens 2, lighting should be even and sufficiently bright so that a participant can see without effort but not so bright that a participant has difficulty looking into the environment. To compensate for the darkness of the visor, dim lights reflecting in the direction of the participant's head are deemed most effective because a tinted visor may cause a loss of contrast in the
physical environment.

\noindent{\textbf{\emph{Experiment duration.}}} 
How long will an experiment run with the same repetitive task? 
It should be long enough for the participant to feel involved, and we can collect sufficient data on \emph{reaction time}. 
It should be short enough that the participant does not feel tired or disengaged, impacting the accuracy of their interactions. 
We tested for a duration of 2 to 10 minutes. 
At 2 minutes, the participants reported that they could not get acquainted with the environment.  
At 10 minutes, the participants reported feeling disinterested after a while. 
We also wanted to test the recovery time of the presence or reaction time (discussed in detail in ~\ref{sec:eval_response_delay}). 
We picked 5 minutes per experiment as it allowed us to obtain at least 60 reaction time readings per experiment, and the participant did not feel tired. 

\noindent{\textbf{\emph{Wait time between blocks and experiments.}}} 
After each block of an experiment, the participant was asked to complete the questionnaires. 
We explored 0-20 minutes of wait times after filling out the questionnaire and before starting the next block. Participants reported 5-minute wait times as sufficient, but a longer break was available upon request. In the main study, we explicitly asked participants if they needed more downtime before each experiment.

\noindent{\textbf{\emph{Experiment and block order.}}} 
For an experiment, we could expose the participant to a control block and then to manipulated trails, or vice versa. 
In either case, there is a risk of obtaining better performance due to greater experience.
\new{Out of an abundance of caution, we decided to run all participants with the control block first and manipulated block second so that any underlying tendency for performance to improve over time would work against our hypotheses.  However, in hindsight, we recognize that counterbalancing would have been a more appropriate way to control such potential effects.}

\noindent{\textbf{\emph{Virtual object placement.}}}
We use Fitts's law~\cite{fittslaw} to calculate the expected time of motor movement for several different positions in the scene. We placed the virtual object parallel to a sitting participant's eye level. 
The object was placed at a $45$ centimeter distance, as recommended by Hololens 2 intractable object guidelines with air tap interaction~\cite{interactable-object}. 
The object's size was tested between $1.4\times 1.4cm$ and $3.5 \times 3.5cm$. 
Both participants felt comfortable with $1.4\times 1.4cm$. 

\noindent{\textbf{\emph{Event and task period}.}} 
Our experiments involve repeating trials in each block. However, repeating events too quickly can make the task more difficult and decrease accuracy, while long breaks between events can break the presence~\cite{break-in-presence}. To find the appropriate interval, we tested intervals from $1-15$ seconds ($s$) but found that periods shorter than $3s$ were too short and caused confusion, while periods longer than $8s$ were boring for participants. Therefore, we settled on a $5s$ interval between cue onsets to balance user comfort and task accuracy.

\noindent{\textbf{\emph{Cue appearance.}}}
We initially tested using a color change as a visual cue for the participant's interaction in both sets of experiments (see details in~\S\ref{sec:experiment}), but it was found to be distracting. We then tried other cues and settled on using objects appearing in the scene to initiate an air tap gesture in the fruit ninja game in realistic and abstract scenarios. A glowing button prompt was tested but found distracting for the coffee mug experiment. The cue was ultimately changed to filling the coffee in the mug, which participants found more engaging.

\noindent{\textbf{\emph{Proximity of interaction.}}} We tested the air tap interaction at a close and far distance. The participant struggled with the far air tap. This could be because the virtual object was placed near the user, and the far-touch interaction moved the virtual object farther from the participant and created an unnecessary distraction. 
The participant found the near air-tap interaction is more natural, so we kept it as the mode of interaction for all of our experiments.

\noindent\new{
{\textbf{\emph{Accuracy of interaction.}}}
To assess the potential for mistriggers, we conducted experiments with pilot study participants. In this pilot, we asked participants to perform random movements with their hands for 10 minutes while avoiding the air tap gesture. Throughout the pilot, our system did not register any unintended interactions by the participants, indicating a low-to-no chance of mistriggers.
}

\new{We took several steps to mitigate the possibility of erroneous gesture detection.  First, before the experiment, we coached participants on performing a successful single air tap gesture.  Second, we did not record any timing data for a trial where an air tap was undetected. Our data shows an average of 50 detected air taps out of 60 possible during the 5-minute interaction, as indicated in table~\ref{table:empirical-exp1} and table~\ref{table:empirical-exp2}. Third, if a participant tapped twice in response to a single cue, we considered only the first measure as the response time.}

\noindent{\textbf{\emph{Feedback.}}} 
\label{sec:pilot-feedback} 
In the realistic vs. abstract experiment, we employed virtual objects disappearing as feedback for successful interaction, which participants found acceptable, similar to the original fruit ninja game. For the plausible vs. implausible experiment, we initially used a subtle coffee-poured sound as feedback, but one participant found it distracting. We then changed to using the mug color change as feedback, which was found to be more subtle and less distracting. Interestingly, participants preferred the ``change in color'' being used as feedback rather than as a prompt for a gesture.

\subsection{Procedure}
Figure ~\ref{fig:study-flow} shows the complete timeline of the user study. 
Upon arrival in the study room, the experimenter welcomed the participants and provided them with verbal introductions and instructions.
Participants subsequently read and signed the IRB-approved informed consent form and completed a pre-questionnaire about demographic characteristics (age, self-identified sex, and familiarity with the MR). 
We also briefed them on MR and HoloLens 2).
We explained the study's flow, the types of experiments, the duration of each block and experiment, their order, and the questionnaires.
We further clarified how to interact with the virtual objects, perform an air tap, and keep their hands in their laps when not air tapping. 
We explained what participants should expect while wearing the headset, how and when to interact with the virtual object, and the cues for air tapping and feedback. 
\new{
After the verbal training, we asked participants to wear the headset and perform five taps on average to select different menu items. The participants manually initiated HoloLens 2's calibration process, which improves visual quality and comfort for the participant. The calibration process is described in HoloLens 2 Manual~\cite{hololens-recalib}.
}

\begin{figure}[t]
    \centering
    \includegraphics[width=0.49\textwidth]{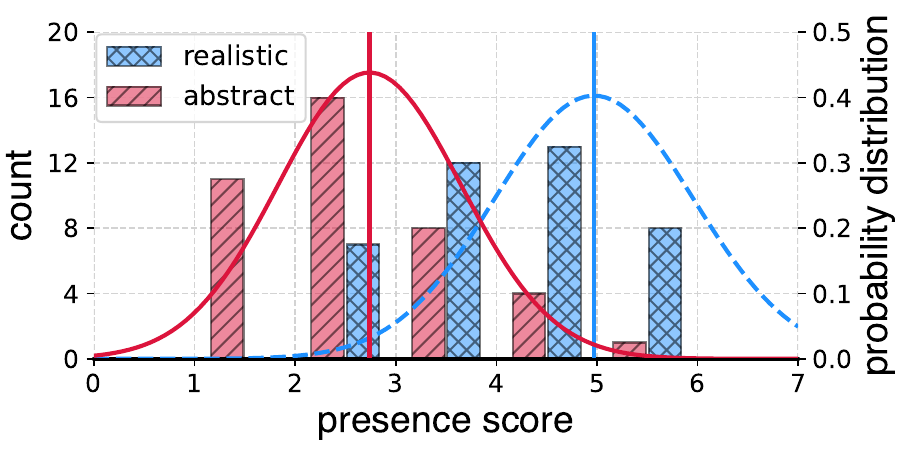} 
    \vspace{-0.65cm}
    \caption{\emph{\textbf{Histogram, means (left y-axis), and fitted Gaussian distribution (right y-axis) using mean questionnaire scores.}}}
    \vspace{-0.25cm}
    \label{fig:user_score_dist_exp1}
\end{figure}

\begin{table}[t]
\small
\begin{center}
\begin{tabular}{|| l | c | c | c | c  ||} \hline
{\bf No} & {\bf realistic} & {\bf abstract} & {\bf t-test} & {\bf \textit{p} }  \\
 & \textbf{($\mu$, $MAD$)} & \textbf{($\mu$, $MAD$)} & & \textbf{($<$0.05)} \\ \hline \hline \hline
ALL & 4.97\textbf{,} 0.90 & 2.74\textbf{,} 0.74 &  19.44  & yes  \\ \hline  \hline
SUS & 4.92, 1.09 & 2.60, 1.89 & 8.35  & yes \\ \hline
IPQ & 4.17, 1.58 & 2.31, 0.57  &  8.98  &  yes \\  \hline
PQ & 5.58, 0.65 & 3.09, 0.99 & 10.03  & yes \\ \cline{2-4}  \hline \hline
REAL & 5.13, 0.54 & 3.04, 0.79 & 17.03  & yes \\ \cline{2-4}  \hline 
REAL, GP, SP & 4.99, 0.71 & 2.94, 0.59 & 6.63  & yes \\ \cline{2-4}  \hline 
ALL - (REAL, GP, SP) & 4.94, 1.40 & 2.94, 1.29 & 1.03  & 0.07 \\ \cline{2-4}  \hline \hline
\end{tabular}
\vspace{-0.25cm}
\caption{\emph{\textbf{Mean, mean absolute deviation (MAD), and paired samples t-test results for all questionnaires.}}}
\label{table:quest-exp1}
\end{center}
\vspace{-1.1cm}
\end{table}

As speeding could potentially override participants' natural instincts, we advised them to complete the task correctly without rushing or stalling. To avoid this instinctual bias, we deliberately kept this information (reaction time measurement) from the participant at the beginning and added this information to the debrief. Participants were also not informed of the frequency of tasks to minimize anticipation. 
We also used the talk-aloud protocol and asked participants to think aloud as they performed the tasks to assess the varied reactions of the participants over time, which the subjective questionnaire does not capture. 
We also monitored the entire process through the first-person user view in Mixed Reality Capture~\cite{mixed-reality-capture}. 
Participants were also advised to stop at any point in the experiment if they felt uncomfortable. 

We world-locked the virtual scene in the environment to ensure that the virtual objects always appeared at the same locations relative to objects in the physical environment rather than at a fixed position relative to Hololen's screen.
The reason for world-locking the scene is to remove the environment variable, and the world-lock is adjusted at the participant's eye level by adjusting the participant's chair. 
We adjusted the headset to the participant’s head and calibrated it to the participant's interpupillary distance for the best visual results. Before the first experiment, for training purposes, participants were asked to interact with the virtual object (an empty dialog box with the next button and cancel) using an air tap. The experimenter and the participant verbally confirmed the success of the first air tap.

We set the order of blocks for the participant as \emph{realistic}, \emph{abstract}, \emph{plausible}, and finally \emph{implausible}. 
In each block, the timer is started only when the participant is comfortable and verbally confirms readiness. At the end of the 5 minutes, the block ends by closing the scene (through the device portal~\cite{device-portal}). 
We then remove the headset from the participant's head and ask them to fill out three post-experience questionnaires (see details in \S\ref{sec:questionnaire}). 
After completing four blocks and the post-experience questionnaires, we thanked the participants for their participation and provided their compensation. The participants were then briefed on the study.
After informing the participants about the purpose of the study, at this stage, we provided them with the option to withdraw from the study. 

\subsection{Hypothesis}

Based on the related work, we developed the following hypothesis for our experiment sets.

\noindent \textbf{H1}: Manipulating the place illusion (virtual object appearance) or plausibility illusion (virtual object behavior) leads to a change in presence. 
    
\noindent \textbf{H2}: Change in presence for a participant leads to a change in the participant's reaction time. 

\noindent \textbf{H3:} Presence and reaction time are correlated.

\section{Results}
\label{sec:eval}
This section presents the results quantifying \emph{presence} using questionnaires and user \emph{reaction time}. 

\begin{figure}[t]
    \centering
    \includegraphics[width=0.49\textwidth]{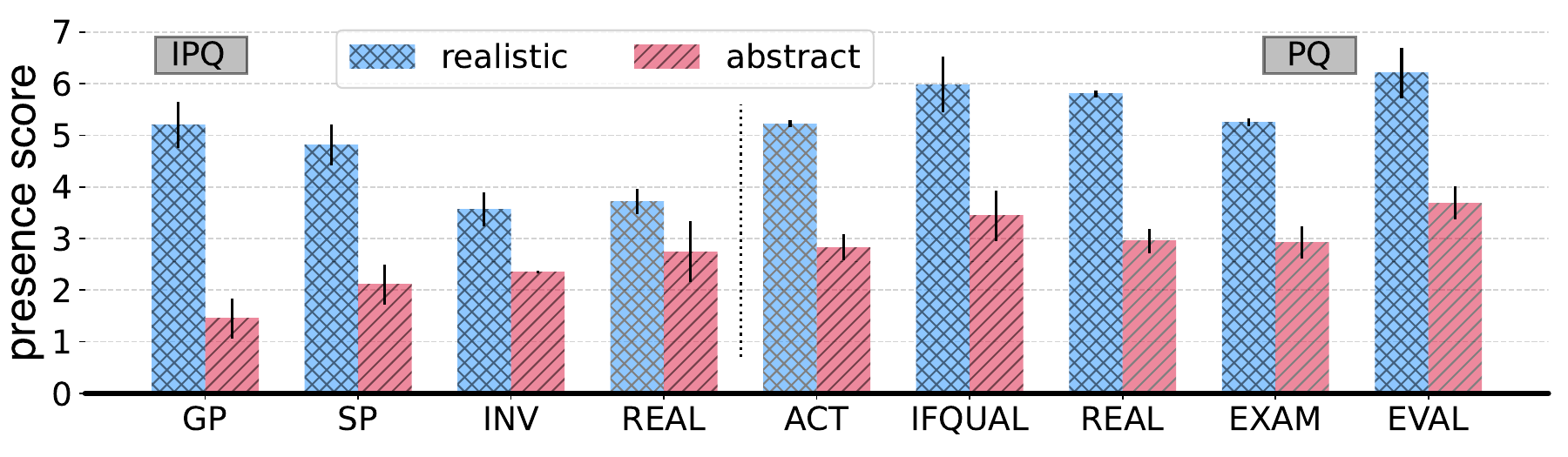} 
    \vspace{-0.65cm}
    \caption{\emph{\textbf{Subscale Scores. IPQ: general presence (GP), spatial presence (SP), involvement (INV), and realism (REAL); PQ: possibility to act (ACT), interface quality (IFQUAL), realism (REAL), possibility to examine (EXAM), and self-evaluation of performance (EVAL).}}}
    \vspace{-0.25cm}
    \label{fig:quest_subscales_exp1}
\end{figure}

\subsection{Experiment Set 1: Realistic vs. Abstract}
\label{sec:exp-set-1-results}
In this experiment, we changed the feelings of \emph{presence} by manipulating the appearance of the object (i.e., \emph{place illusion}). We get presence scores from the questionnaires and reaction time scores from HoloLens. 

\begin{table}[t]
\small
\begin{center}
\begin{tabular}{|| l | c | c | c ||} \hline
{\bf Experiment} & {\bf Reaction Time (ms)} & \textbf{Change} &  \textbf{No. of}  \\
 & \textbf{($\mu$ | $MAD$)} & \textbf{(\%)} & \textbf{Interactions}\\ \hline \hline \hline
realistic & 954 | 170 & & 50 \\ \hline 
abstract & 1313 | 290 & 37.63 & 44\\ \hline \hline
\end{tabular}
\vspace{-0.2cm}
\caption{\emph{\textbf{Average, MAD, \%age change of user reaction times, and average number of interactions across the two sets of experiments.}}
\label{table:empirical-exp1}}
\end{center}
\vspace{-1.1cm}
\end{table}

\vspace{-0.1cm}
\subsubsection{Presence Questionnaire Scores}
\label{sec:eval_questionnaire_exp1}
\vspace{-0.05cm}
We collect answers to all questionnaires on a 7-point scale.
The \emph{presence} score for all the questionnaires, individually or combined, is computed as the average of 7-point scores. 

\noindent{\textbf{Questionnaire scores.}}
Our presence scores are collected from the same set of participants under two different conditions: realistic vs. abstract.  
We use a paired samples t-test, with a \emph{null hypothesis} that \emph{mean of two sets of experiments is equal}, to determine if the presence scores changed between realistic and abstract experiments. 
Before applying the t-test, we verified the normality of the difference between the presence scores for the two experiments.  
The results of this experiment are reported in Table~\ref{table:quest-exp1}. The difference in presence score between realistic experiment (M = 4.97; MAD = 0.90) and abstract experiment (M = 2.74; MAD = 0.74) was significant (t (40) = 12.85; p < 1.32e-15). Therefore, we can reject the null hypothesis and state that the presence of subjects changed across experiments. Figure~\ref{fig:user_score_dist_exp1} shows the histogram of the scores across users and its probability distribution.

\noindent{\textbf{Subscales.}}
While our aggregate results demonstrate that the presence score changed as we altered the realism of the objects, we want to investigate the factors that contributed to the change in presence.  
The mean scores for all subscales and the aggregate realism scores across questionnaires are shown in Figure~\ref{fig:quest_subscales_exp1}. 
First, the realism questions constitute 11 of the 39 questions and show a significant change in the presence scores. 
This suggests that the realism significantly changed across the two experiments, also confirmed by the t-test. 
We also report the disaggregated presence scores for realism- and presence-related questions to analyze how factors other than realism and feeling of presence affect the overall presence score.
While the null hypothesis is true, the p-value is very small, indicating that other aspects had a lesser impact but require further investigation.

\begin{figure}[t]
    \centering
    \includegraphics[width=\linewidth]{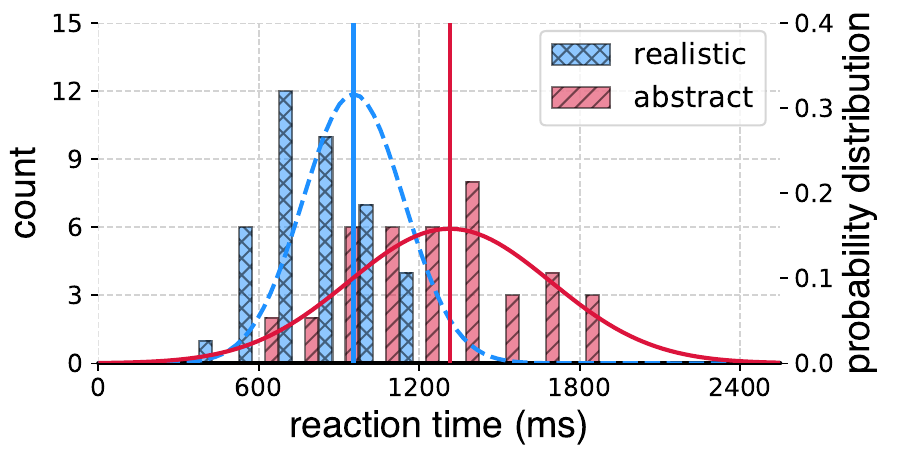} 
    \vspace{-0.55cm}
    \caption{\emph{\textbf{Histogram, means (left y-axis), and fitted Gaussian distribution (right y-axis) using the average reaction time across participants.}}}
    \vspace{-0.25cm}
    \label{fig:user_reaction_time_dist_exp1}
\end{figure}

\begin{figure}[t]
    \centering
    \includegraphics[width=\linewidth]{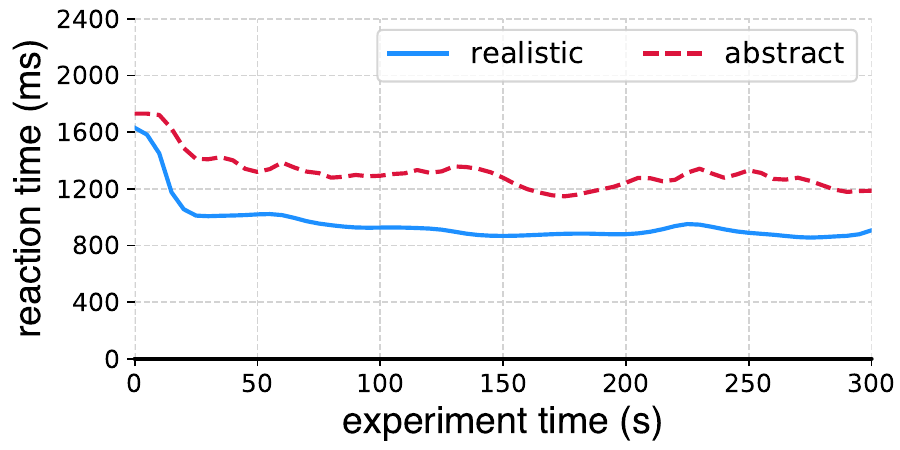} 
    \vspace{-0.55cm}
    \caption{\emph{\textbf{Average user \emph{reaction time} for different experimental settings. User reaction time recovers (decreases) over time.}}}
    \vspace{-0.65cm}
    \label{fig:recovery_exp1}
\end{figure}

\noindent{\textbf{User characteristic and presence.}}
We performed a regression analysis using F-test\footnote{$F(X, Y)$: $X$, $Y$ are degrees of freedom between and within groups, respectively. $X=$ total groups$-1$, $X=$ group size$-$total groups.} to check if age, gender, and familiarity with MR had any impact on \emph{presence} scores. 
In this analysis, the null hypothesis is that a regression model based on a given variable is not a better fit than a simple intercept-only model. 
We observed that age, gender, and familiarity with MR did not have any effect on overall \emph{presence} scores, as we obtained values of $F(1,38) = 0.96, p = 0.34$,  $F(1,38) = 1.64, p = 0.96$, and  $F(1,38) = 2.19, p = 0.71$, respectively.

\subsubsection{Reaction Time}
\label{sec:eval_response_delay}

To evaluate participants' reaction time, we record a time-stamp when the cue appears and a time-stamp when the user's action is recorded. We use the difference between these two timestamps as the reaction time.
\new{
There are instances of ``no-triggers'' where either the participant does not respond, or the air tap does not register. 
We remove the corresponding cue from our observations if the air tap is not registered. As a result, our experiments recorded, on average, 50 responses out of the maximum possible 55-60, indicating that around 10-15\% of the time, participants did not respond, or the air tap was not recorded. Removing this data ensures that ``no-triggers'' do not impact our findings. 
}
Table~\ref{table:empirical-exp1} presents the high-level results of the experiments.

\noindent{\textbf{Reaction time values.}} 
Figure~\ref{fig:user_reaction_time_dist_exp1} shows the distribution of average reaction time scores across the users. 
Similar to the presence scores, we used a t-test to evaluate the null hypothesis that the median reaction time across the two experiments is equal.
Our results show a significant difference in average reaction times between the appearance conditions:  954ms in the realistic condition and 1313ms in the abstract condition with t-statistics of t(40) = 8.71 and $p < 1.09e^{-16}$.
This rejects the null hypothesis and also presents a significant difference of 37.63\%.

\noindent{\textbf{Reaction time recovery.}} 
Next, we examined the reaction time of users over time. In Figure~\ref{fig:recovery_exp1}, we observe that participants took significantly longer to respond to the cues at the start of the experiment. The reaction time drastically dropped and settled to a steady-state within the first 30 seconds. After that, there was only a modest improvement in the reaction time over the duration of the experiment. 

\noindent{\textbf{Additional statistical analyses.}}
We also measured the number of times a user was able to respond to the cue. 
Since the number of cues between experiments differed due to slight variations in the experiment duration, we only considered the first 60 cues to collect these statistics. 
We see that users could complete more interactions in the realistic appearance condition than in the abstract appearance condition.

\begin{table}[t]
\small
\begin{center}
\begin{tabular}{|| l | c | c | c | c ||} \hline
{\bf No} & {\bf plausible} & {\bf implausible} & {\bf t-test} & {\bf \textit{p} }  \\
 & \textbf{($\mu$, $MAD$)} & \textbf{($\mu$, $MAD$)} & & \textbf{($<$0.05)} \\ \hline \hline \hline
ALL &  5.17\textbf{,} 0.93 & 2.81\textbf{,} 0.91 &  8.05  & yes  \\ \hline  \hline
SUS &  5.34, 1.09 & 2.70, 1.13 &  7.52  & yes \\ \hline
IPQ &  4.51, 1.35 & 2.63, 0.90  &  6.10  & yes \\  \hline
PQ & 5.61, 0.79  & 2.98, 1.30 &  7.88  & yes \\ \cline{2-4}  \hline \hline
Plausible & 4.73, 0.89 & 2.77, 0.57 & 5.23  & yes \\ \cline{2-4}  \hline 
Plaus., GP, SP & 4.94, 0.43 & 2.62, 0.77 & 9.05  & yes \\ \cline{2-4}  \hline 
ALL - (Plaus., GP, SP) & 5.36, 1.34 & 2.97, 1.16 & 4.18  & 0.06 \\ \cline{2-4}  \hline \hline
\end{tabular}
\vspace{-0.25cm}
\caption{\emph{\textbf{Mean, MAD, and t-test results for all questionnaires.}}}
\label{table:quest-exp2}
\end{center}
\vspace{-0.79cm}
\end{table}

\begin{figure}[t]
    \centering
    \includegraphics[width=\linewidth]{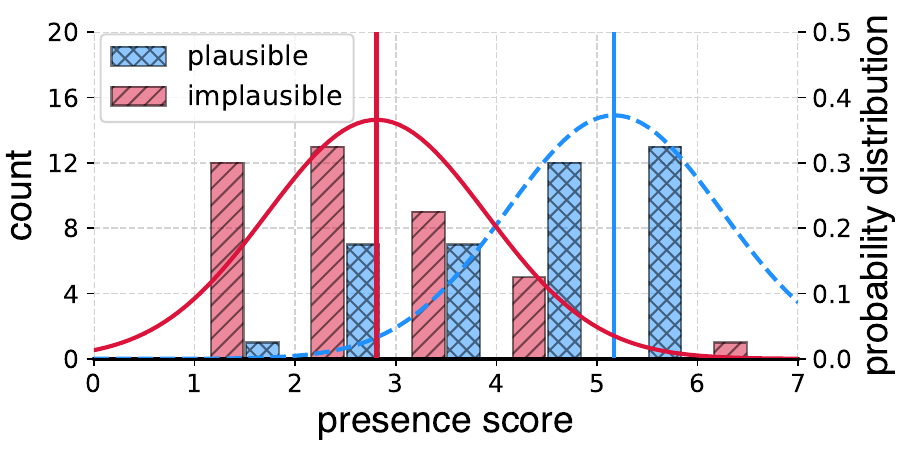} 
    \vspace{-0.7cm}
    \caption{\emph{\textbf{Histogram, means (left y-axis), and fitted Gaussian distribution (right y-axis) using mean questionnaire scores.}}}
    \vspace{-0.55cm}
    \label{fig:user_score_dist_exp2}
\end{figure}

\subsubsection{Experiment Set 1: Discussion} 

Our results suggest that the place illusion part of our first hypothesis, ``\emph{H1: manipulating the place illusion (appearance of a virtual object) leads to change in presence}'', is valid. 
We successfully altered the presence by manipulating the object's appearance from realistic to abstract. 
Prior work on manipulating the place illusion to alter the presence also supports our results~\cite{break-in-presence}. 
Additionally, the high correlation between questionnaire scores between conditions suggests that the users who felt a greater presence in one condition also felt a lower presence in the second condition. 
Our second hypothesis, ``\emph{H2: Change in presence for a participant leads to change in participant’s reaction time}'' is also valid. Our first hypothesis confirms that the presence changed from the experiment with a realistic object to the experiment with an abstract object. Simultaneously, we observed a significant increase in the user reaction time as participants moved from realistic to abstract object experiments.
Finally, the recovery time analysis also yields interesting points. The reaction time for the manipulated experiment block is always higher than the control experiment block. 
Furthermore, the reaction time is steady after the initial few seconds. 
This means the participants initially took some time to get acquainted with the environment, leading to an increase in their reaction time. 
As time went on, they interacted with the virtual objects in a fairly consistent manner. 

\begin{figure}[t]
    \centering
    \includegraphics[width=\linewidth]{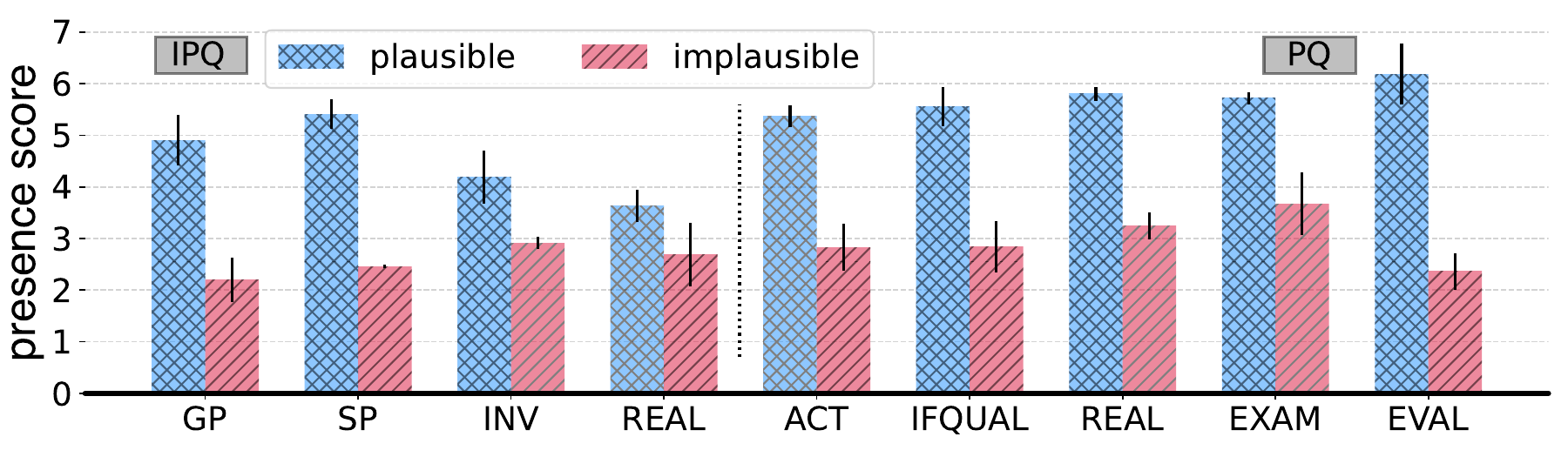} 
    \vspace{-0.75cm}
    \caption{\emph{\textbf{Subscale Scores. See Figure~\ref{table:quest-exp2} caption for subscale acronyms.}}}
    \vspace{-0.25cm}
    \label{fig:quest_subscales_exp2}
\end{figure}

\subsection{Experiment Set 2: Plausible vs. Implausible}
\label{sec:exp-set-1-results}

As in the previous experiment set, we get presence scores from questionnaires and reaction times from HoloLens 2. 

\subsubsection{Presence Questionnaire Scores}
\label{sec:eval_questionnaire_exp2}
In this experiment, we changed the feelings of \emph{presence} by manipulating the behavior of the object (i.e., \emph{plausibility illusion}). The experiment and analysis setup is the same as the previous experiment set. 

\noindent{\textbf{Questionnaire scores.}}
Our presence scores are collected from the same set of participants under two different conditions: plausible vs. implausible.  
We use a paired samples t-test, with a \emph{null hypothesis} that \emph{mean of two sets of experiments is equal}, to determine if the presence scores changed between plausible and implausible experiments. 
Before applying the t-test, we verified the normality of the difference between the presence scores for the two experiments.  
The results of this experiment are reported in Table~\ref{table:quest-exp2}. 
The difference in presence score between plausible experiment (M = 5.17; MAD = 0.93) and implausible experiment (M = 2.81; MAD = 0.91) was significant (t (40) = 8.05; p < 8.11e-10).
Therefore, we can reject the null hypothesis and state that the presence of subjects changed across experiments. 
Figure~\ref{fig:user_score_dist_exp2} shows the histogram of the scores across users and its probability distribution.

\noindent{\textbf{Subscales.}}
While our aggregate results demonstrate that the presence score changed as we altered the plausibility of the objects, we want to investigate the factors that contributed to the change in presence. The mean scores for all subscales and the aggregate plausibility scores across questionnaires are shown in Figure~\ref{fig:quest_subscales_exp1}. 
First, we filtered the questions relating to plausibility, comprising 8 out of 39 questions, which show a significant change in the presence scores. This suggests that plausibility significantly changed across the two experiments, also confirmed by the t-test. We also report the disaggregated presence scores for plausible- and presence-related questions and the rest of the questions to analyze how factors other than plausibility and feeling of presence affect the overall presence score. While the null hypothesis is true, the p-value is very small, indicating other aspects had a lesser impact but require further investigation.

\begin{table}[t]
\small
\begin{center}
\begin{tabular}{|| l | c | c | c ||} \hline
{\bf Experiment} & {\bf Reaction Time (ms)} & \textbf{Change} &  \textbf{No. of}  \\
 & \textbf{($\mu$ | $MAD$)} & \textbf{(\%)} & \textbf{Interactions}\\ \hline \hline \hline
plausible & 930 | 240 & & 54 \\ \hline  
implausible & 1182 | 210 & 27.10 & 51\\ \hline \hline
\end{tabular}
\vspace{-0.2cm}
\caption{\emph{\textbf{Average, MAD, \%age change of user reaction times, and average number of interactions across the two sets of experiments.}}}
\label{table:empirical-exp2}
\end{center}
\vspace{-0.99cm}
\end{table}

\subsubsection{Reaction Time}
\label{sec:eval_response_delay_exp2}
We collected 50 data points per user per experiment on average\new{ after post-processing data as described in Section~\ref{sec:eval_response_delay} for the first experiment.} 
Table~\ref{table:empirical-exp2} presents the results of the experiments.

\begin{figure}[t]
    \centering
    \includegraphics[width=\linewidth]{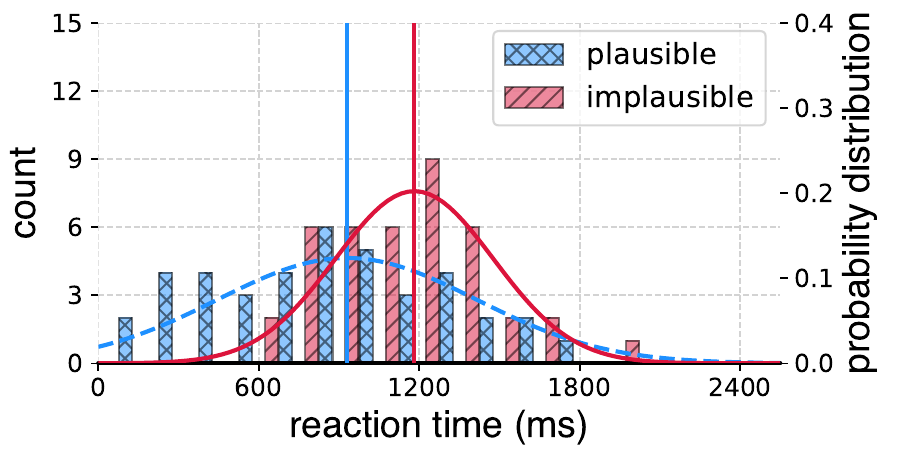} 
    \vspace{-0.6cm}
    \caption{\emph{\textbf{Histogram, means (left y-axis), and fitted Gaussian distribution (right y-axis) using the average reaction time across participants.}}}
    \vspace{-0.20cm}
    \label{fig:user_reaction_time_dist_exp2}
\end{figure}

\noindent{\textbf{Reaction time values.}} 
Figure~\ref{fig:user_reaction_time_dist_exp2} shows the distribution of average reaction time scores across the users. Similar to the presence scores, we used a t-test to evaluate the null hypothesis that the median reaction time across the two experiments is equal. Our results show a significant difference in average reaction times between the two plausibility conditions:  930ms in the plausible condition and 1182ms in the abstract condition with t-statistics of t(40) = 11.67, and $p < 1.93e^{-13}$.
This rejects the null hypothesis and also presents a significant difference of 27.10\%.
As in the first experiment set, we see decreased interactions after the manipulation. 

\begin{figure}[t]
    \centering
    \includegraphics[width=\linewidth]{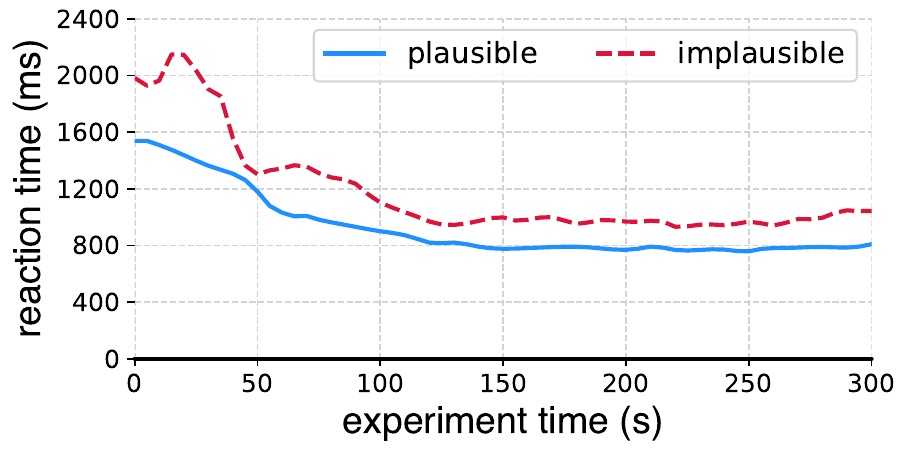} 
    \vspace{-0.65cm}
    \caption{\emph{\textbf{Average user \emph{reaction time} for different experimental settings. User reaction time recovers (decreases) over time.}}}
    \vspace{-0.75cm}
    \label{fig:recovery_exp2}
\end{figure}

\noindent{\textbf{Reaction time recovery.}} 
Next, we examined the reaction time of users over time. In Figure~\ref{fig:recovery_exp2}, participants took slightly longer than 60 seconds to reach a steady state for the plausible vs. implausible experiment, which was longer than the first experiment set.

\begin{figure*}[t]
    \centering
    \begin{tabular}{cccc}
    \includegraphics[width=0.23\textwidth]{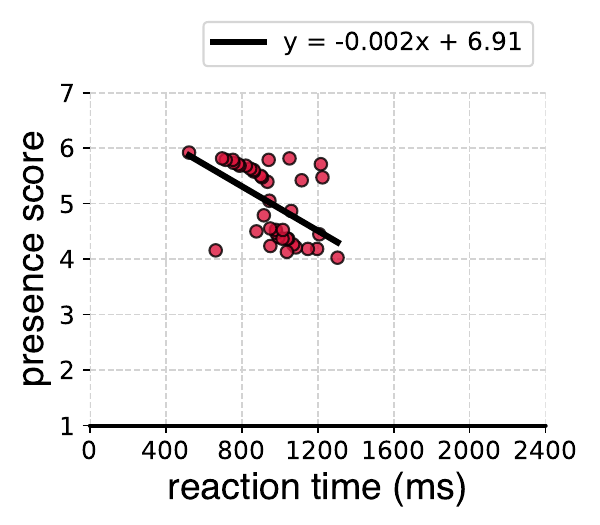} &
    \includegraphics[width=0.23\textwidth]{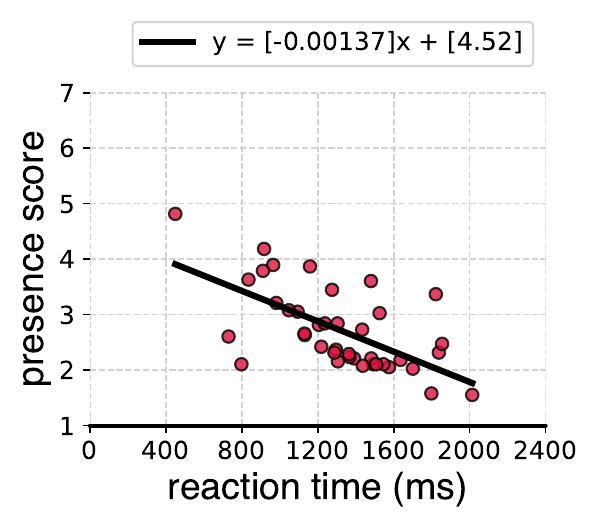} &
    \includegraphics[width=0.23\textwidth]{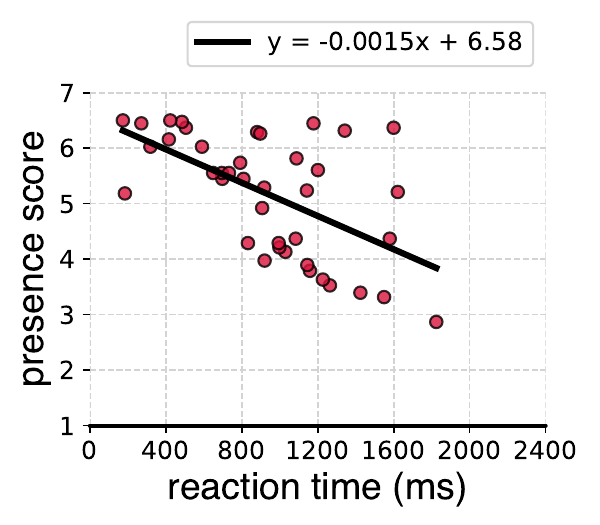} &
    \includegraphics[width=0.23\textwidth]{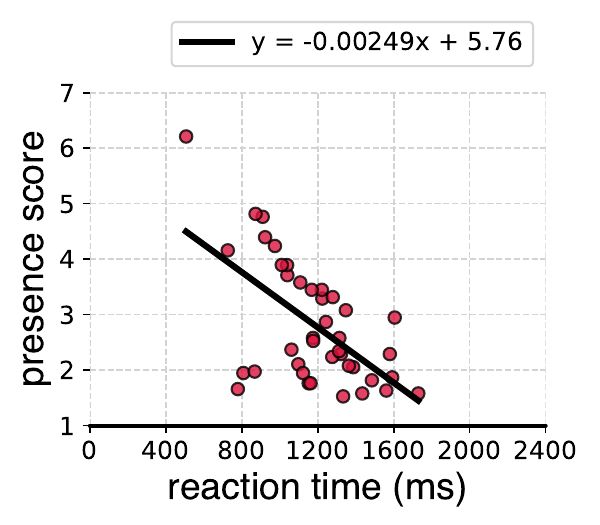} \vspace{-0.05cm} \\
    (a) realistic & (b) abstract & (c) plausible & (d) implausible \\
    \end{tabular}
    \vspace{-0.3cm}
    \caption{\emph{\textbf{Presence vs. Reaction Time: Presence decreases as reaction time increases. Reaction time and presence also show a modest correlation: overall (-0.65), realistic (-0.51), abstract (-0.63), plausible (-0.57), and implausible (-0.59). Each red circle represents a study participant. The black line is the linear regression fit for the data.}}}
    \vspace{-0.5cm}
    \label{fig:reaction_time_presence}
\end{figure*}

\subsubsection{Experiment Set 2: Discussion}

This experiment aimed to establish that presence can be modified in ways other than manipulations to the place illusion done in the previous experiment. For this experiment, the plausibility illusion part of our hypothesis ``\emph{H1: manipulating the plausibility illusion (behavior of virtual object) leads to change in presence}'' is valid. We altered the presence by manipulating the object's behavior from plausible to implausible. Gravity is a crucial aspect of our lives, and we expect objects to behave in specific ways under gravity. If their behavior is not plausible according to the laws of physics, it leads to degraded presence. Similar to the previous experiment, we saw a high correlation between questionnaire scores, suggesting that plausibility illusion was the factor in altering presence. The second hypothesis, H2, can be accepted based on our results. The only difference is that the change in presence was due to the change in plausibility illusion and not the place illusion. The recovery time analysis for this experiment set yields similar results as the previous experiment. However, users took longer to reach the steady state, and reaction times varied over time. This means that manipulating the appearance of the objects has less effect on participants than behavior manipulation. This is understandable as humans are more likely to see objects with non-standard appearance as opposed to observing objects that do not conform to gravity. 

\section{Discussion}
\label{sec:disc}
In this section, we discuss the implications of the quantitative results presented in the previous section. 

Our first hypothesis, ``\emph{H1: Manipulating the place illusion (appearance of a virtual object) leads to change in the presence}'' for the first experiment set can be accepted. The presence score of participants for the manipulated experiment was more than 2 points lower on a 7-point scale than the control experiments. As we used everyday virtual objects, bananas in this case, a change in their texture significantly alters our strong experience-based prior.
The results from the second experiment set further support this statement and result in the acceptance of H1. 
As all the participants experience gravity at all times, an object defying gravity is implausible and causes a break in presence. 
This statement is supported by prior work asserting that break-in plausibility, gravity-defying behavior, leads to a change in presence~\cite{break-in-presence}.
Finally, our hypothesis is acceptable for the individual questionnaires and their subscales, as the change in presence is the same as the overall results. 

Our user \emph{reaction time} results suggest that our second hypothesis ``\emph{H2: Change in presence for a participant leads to change in participant's reaction time}'' can be accepted.
As the users became more familiar with the environment and setup, the reaction times from one set of experiments to the other decreased. 
We also observed an improvement in the number of interactions across the two sets.
However, this improvement could not overcome the increase in reaction times due to our manipulations within a set of experiments. 
Within a block, across both experiments, the reaction time drastically improved at the start but quickly reached a steady state. 
The difference in reaction time between the blocks of an experiment remained constant over time. 
This indicates that the recovery effect is consistent across blocks, and our manipulations influenced the change in reaction times of the participants.

\subsection{Presence - Reaction Time Correlation} 
Up to this point, we have established that we were able to change the presence, and participants' reaction times changed as their presence changed. 
Figure~\ref{fig:reaction_time_presence} shows the scatter plots of participants' \emph{presence} scores and their \emph{reaction times} for all the experiments. 
We observe that \emph{reaction time} has an inverse correlation with \emph{presence} score, albeit with different slopes for the linear regression model across experiments. 
Our results highlight that appearance and behavior manipulations impacted the presence questionnaire scores and reaction time values similarly when aggregated across users. We also observed a moderate correlation between the presence scores and reaction times of individual users.

This is a significant result and suggests that \emph{presence} impacts \emph{reaction time}, at least within the confines of our study. This means we can accept our hypothesis that ``\emph{H3: presence and reaction times are correlated}''. 
However, the correlation values lie in the modest range between -0.5 and -0.65. 
This means that further analysis is needed to fully establish that reaction time can be used to quantify presence. 
This is because a given change in \emph{presence} score or \emph{reaction time} does not elicit the same change in the other variable across all experiments. 
Also, the individual differences between different experiments are significant enough to warrant that the use of this relationship should be scenario-specific. 
Having said that, we can at least state that if the presence of a user increases or decreases, the user reaction time will also decrease or increase respectively.

\color{black}
\subsection{Modeling Reaction Time--Presence Relationship}
Our results in the previous section demonstrate a high correlation between presence and reaction time. We leverage this correlation to build a classification model that takes the user's reaction time as input and outputs the level of presence for the user. We modeled the problem as a classification task, as the presence is measured in discrete states. Next, we describe the experimental setup and the training process. 

\begin{figure}[t]
    \centering    \includegraphics[width=\linewidth]{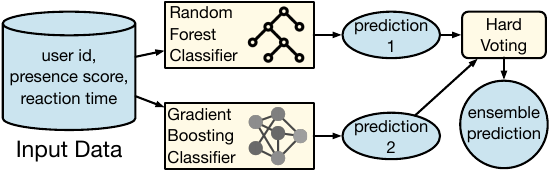} 
    \vspace{-0.3cm}
    \caption{\new{\emph{\textbf{The architecture of reaction time--to--presence model.}}}}
    \vspace{-0.7cm}
    \label{fig:model-arch}
\end{figure}

\subsubsection{Experimental Setup} 

\noindent\textbf{Models.} To perform the classification task, we utilize an ensemble machine learning model incorporating a hard voting classifier~\cite{voting-classifier}. The model integrates a random forest classifier~\cite{rf-classifier} and a gradient boost classifier~\cite{gb-classifier}. Figure~\ref{fig:model-arch} shows the overall architecture of the classification model. We also tested several other classification models, such as K-nearest neighbors, support vector machines, and naive Bayes. However, they did not perform better than the ensemble model we chose for the classification task.

\vspace{0.08cm}
\noindent\textbf{Dataset.} Our dataset consists of reaction time and presence values for 40 users. For each user, we have reaction times and presence ratings under 4 conditions: realistic, abstract, plausible, and implausible. In total, our dataset contains 160 data points corresponding to the 4 conditions experienced by 40 users.

\noindent\textbf{Training.} We use an 80-20 dataset split for training and testing, respectively. We perform the split at the user-level data, meaning that all the data for a given user is used in training or testing. As a result, we use data from 32 randomly selected users for training and 8 for testing. We use grid search with 5-fold cross-validation for training the individual classifiers with the parameters specified in Table~\ref{table:model-training}. Finally, we use a voting classifier with a hard voting configuration to choose between the two classifiers. 

\noindent\textbf{Output variable.} 
Our presence questionnaire produced ratings of presence on a continuous scale from 0 to 7. However, it does not necessarily follow that we should design our model to estimate 7 discrete levels of presence. In many applications, knowing if presence is high or low could be sufficient. For this reason, we evaluated the efficacy of our model in estimating presence using either 2, 3, or 7 partitions of the continuous data.  When using 2 levels, scores below 3.5 were assigned to level 1 and scores of 3.5 or above to level 2. When using 3 levels, scores between 0-2.33 were assigned to level 1, 2.34-4.66 to level 2, and 4.67-7 to level 3.

\begin{table}[t]
\color{black}
\begin{center}
\begin{tabular}{|| l | c ||} \hline
{\bf Random Forest Parameters} & {\bf Values}  \\ \hline \hline \hline
no. of estimators & [50, 100, 150, 200, 500] \\ \hline  
maximum depth & [None, 5, 10, 15, 20]  \\ \hline 
min. samples split & [2, 5, 10]  \\ \hline 
min. samples leaf & [1, 2, 4]  \\ \hline \hline
{\bf Gradient Boosting Parameters} & {\bf Values}  \\ \hline \hline \hline
no. of estimators & [50, 100, 150, 200, 500] \\ \hline  
maximum depth & [3, 5, 7]  \\ \hline 
learning rate & [0.01, 0.1, 0.2]  \\ \hline \hline
\end{tabular}
\vspace{-0.2cm}
\caption{\color{black} \emph{\textbf{Parameters for the grid search with cross-validation.}}}
\label{table:model-training}
\end{center}
\vspace{-0.85cm}
\end{table}

\subsubsection{Model Evaluation}
The accuracy of our model depends on two factors: the number of presence classes and the amount of data used for model training. Next, we evaluate the effect of these parameters on the model's accuracy.

\begin{figure}[t]
    \begin{tabular}{cc}
    \centering
    \includegraphics[width=0.46\linewidth]{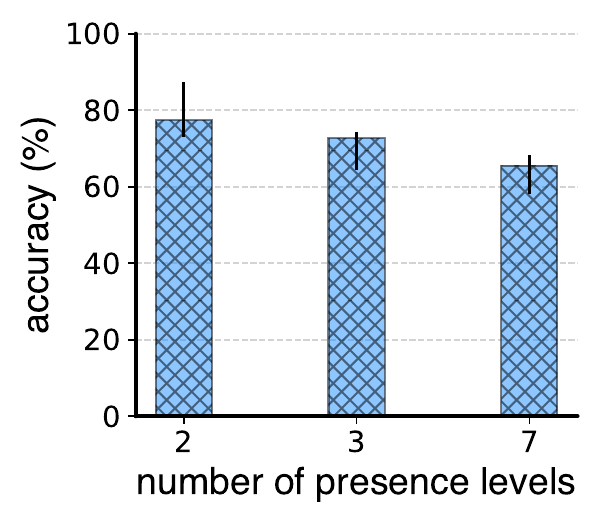} &
    \includegraphics[width=0.46\linewidth]{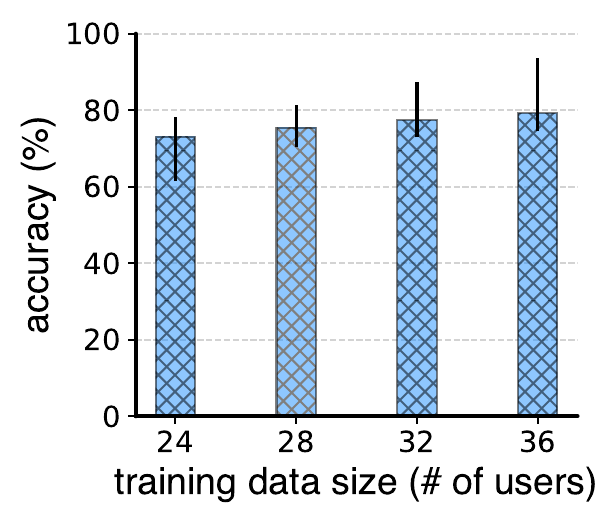} \vspace{-0.05cm} \\
    \new{(a) Effect of Presence Levels} & \new{(b) Effect of Training Size}\\
    \end{tabular}
    \vspace{-0.3cm}
    \caption{\emph{\new{\textbf{The effect of the number of presence levels (a) and the training data size (b) on the model's accuracy. The error bars represent the min and max values observed across 15 experiment runs.}}}}
    \vspace{-0.6cm}
    \label{fig:model-eval}
\end{figure}

\noindent\textbf{Effect of Number of Presence Classes.} Figure~\ref{fig:model-eval}(a) shows the effect of number of presence levels (on $x$-axis) and model's accuracy in percentage (on $y$-axis). We use 80\% of the data, 32 users, for training and 20\% of the data, eight users, for testing. We observe that the accuracy decreases as the number of presence classes increases. However, our model shows good accuracy despite the small data set of 128 observations. The accuracy is significantly higher than a purely random predictor with 1/(no. of presence levels), e.g., 50\% for two levels, 33\% for three levels, and 14.28\% for seven levels. 

\noindent\textbf{Effect of Training Data Size.} Figure~\ref{fig:model-eval}(b) demonstrates the effect of training data size (on $x$-axis) on classification accuracy (on $y$-axis). In this experiment, we set the number of classes to two. We observe that model accuracy improves as the training data size increases. However, even with a very small training data size, the model performs quite well and achieves an accuracy of 73.09\%. The accuracy improves to 79\% when the number of users increases to 36. The upward trend shows that more data can further improve the model performance. 

\noindent\emph{\textbf{Key Takeaway.} Our ensemble classification model estimates presence levels using the reaction time values with high accuracy, which depends on the number of presence levels and training data size. However, the accuracy can be further improved by using data from more users.}

\color{black}

\section{Limitations and Future Work}
\label{sec:future}
For this study, we used two scenes consisting of simple one-object scenarios, a design choice to minimize the effect of variables other than presence on the reaction time. 
\new{
This study establishes the relationship between reaction time and presence when presence is altered by changing the realism and plausibility of virtual objects. While we do not have any indications that suggest this relationship will not hold if its presence is altered by any other method in a more complex environment, there is a possibility that the setup may not be sensitive to the broader effects of varying the feelings of presence on reaction time in a multi-object virtual scene. In future work, we plan to experimentally investigate how presence is affected by factors other than realism and plausibility and how it relates to reaction time with different degrees of scene complexity, cognitive load, and dynamic physical environment.
}

In this study, we investigated the effects of varying the feelings of presence with only periodic tasks and active interaction (air tap). It's worth investigating the relationship between presence and reaction time under different conditions, such as with non-periodic tasks, or with different response measures, such as eye gaze.
Participants reported their presence levels using questionnaires that leverage Likert scales. Other questionnaires that rely on different assessment mechanisms, such as open-ended questions, might reveal additional insights.
We also acknowledge that our proposed approach depends on user interactions to measure the reaction time. Our technique may fail to produce any measurement in virtual scenes with little or no interaction. 
Future work should consider using other backup mechanisms like eye-gaze tracking to combat low-interaction scenarios. 

We have not found any effect of gender, age, or familiarity with MR on this study's presence or reaction time. This can be due to our purposefully simple design, but in a complex scenario with some applications, these user characteristics may affect the presence or reaction time. In addition, we have not tested the effect of the break in the presence, but it may impact the presence and reaction time. 
In the future, the break in the presence can be tested with the reaction time as a measure and may add to the discussion of the presence-reaction-time relationship. 
This study did not ask participants to complete a cybersickness questionnaire. However, during the study with participants, we asked participants to report any discomfort they felt during or at the end of the experiments. None of the participants reported any feeling of discomfort. This may be due to the limited exposure time and a wider adjustable interpupillary distance range that Hololens 2 offers ~\cite{hololens2}. However, this variable can be tested in isolation to isolate the potential impact of cybersickness in varying the feelings of presence.

Despite the limitations of the work and the opportunity for improvements, we argue that our results present sufficient evidence of a relationship between \emph{presence} and \emph{reaction time} to justify a further discussion of whether a performance-based metric such as \emph{reaction time} can be used to describe \emph{presence}. 
Post-experience questionnaires are the most commonly used measures of \emph{presence} in previous work. 
However, a significant disadvantage of such questionnaires is that they are based on the subjects' memories of the experience. 
Such memories can reflect an inconsistent and incomplete picture of the experience. 
\emph{Reaction time}, on the other hand, is a passive and objective measure that does not depend on the comprehension of the user and the memory of the experience. 
\revtwo{
In our work, we have developed a preliminary model that maps the reaction time to presence. }
In the future, with additional investigation, this model can serve as a measure of \emph{presence} and as a feedback loop that developers can use to improve the run-time experience since it measures the phenomenon when it is perceived. 

\revtwo{
Additionally, a model can be developed that takes the presence as the input and yields reaction time as the output, which could describe how much of an ill effect a decrease in presence might cause. However, it must be noted that reaction time is an objective metric that can be measured, and presence is a subjective sensation that even people themselves have difficulty reliably quantifying. This is why our paper aims to use reaction time as an alternative and potentially more robust mechanism for assessing the presence, not for predicting the effect of lower presence on reaction time.
}

\section{Conclusion}
\label{sec:conclusion}
\vspace{-0.1cm}
We presented a user study (N=40) to understand the relationship between \emph{presence} and \emph{reaction time}. 
We changed the sense of \emph{presence} of the participants by manipulating \emph{appearance (place illusion)} and the non-task-relevant \emph{behavior (plausibility illusion)} of the virtual object and systematically measured the \emph{reaction time} of the participants in response to visual stimulus. 
Our post-experience questionnaires show a significant change in
the \emph{presence} across experiments. 
Similarly, we see a significant change in user \emph{reaction time} as we vary feelings for \emph{presence}. 
Our analysis shows a negative correlation between the presence and \emph{reaction time}.
Furthermore, our study provides insight into the considerations
for using \emph{reaction time} as a possible measure of \emph{presence}, as well as preliminary recommendations on the possibilities of future research to understand better the relationship between \emph{presence} and \emph{reaction time}. 
We found that as the average \emph{presence} score for the two illusions decreased from 4.97 to 2.74 and 5.17 to 2.81 (on a 7-point scale), the average \emph{reaction time} increased by 37.63\% and 27.10\%, respectively.
\new{We developed a model that estimates a user's presence level using reaction time values with high accuracy of up to 80\%. While our study suggests that reaction time can be used as a measure of presence, further investigation is needed to improve the accuracy of the model.}

\bibliographystyle{abbrv-doi-narrow}
\bibliography{paper}

\end{document}